\documentclass{emulateapj}
\newcommand{\Lyalpha}{Ly$\alpha$}
\newcommand{\Mdot}{\ensuremath{\dot{M}}}
\newcommand{\Mdoteta}{\ensuremath{\Mdot_\eta}}
\newcommand{\Mdotc}{\ensuremath{\Mdot_\mathrm{c}}}
\newcommand{\veta}{\ensuremath{v_\eta}}
\newcommand{\vc}{\ensuremath{v_\mathrm{c}}}
\newcommand{\vt}{\ensuremath{v_\mathrm{t}}}
\newcommand{\vC}{\ensuremath{v_\mathrm{C96}}}
\newcommand{\R}{\ensuremath{\mathcal{R}}}
\newcommand{\rstar}{\ensuremath{r_{\ast}}}
\newcommand{\rc}{\ensuremath{r_\mathrm{c}}}
\newcommand{\Rsol}{\ensuremath{R_{\odot}}}
\newcommand{\Ne}{\ensuremath{n_\mathrm{e}}}
\newcommand{\nc}{\ensuremath{n_C}}
\newcommand{\phic}{\ensuremath{\phi_C}}
\newcommand{\phistar}{\ensuremath{\phi_\star}}
\newcommand{\xpeak}{\ensuremath{x_\mathrm{peak}}}
\newcommand{\Lline}{\ensuremath{L_\mathrm{line}}}
\newcommand{\betalocal}{\ensuremath{\beta_\mathrm{local}}}
\newcommand{\Dsep}{\ensuremath{D_\mathrm{sep}}}

\newcommand{\Msol}{\ensuremath{M_{\odot}}}
\newcommand{\Msolpy}{\ensuremath{\Msol~\mbox{yr}^{-1}}}

\newcommand{\kmps}{\ensuremath{\mbox{km~s}^{-1}}}
\newcommand{\kev}{\ensuremath{\mbox{keV}}}
\newcommand{\angstrom}{\ensuremath{\mbox{\AA}}}
\newcommand{\mA}{\ensuremath{\mbox{m\AA}}}
\newcommand{\pcc}{\ensuremath{\mbox{cm}^{-3}}}
\newcommand{\pcmsq}{\ensuremath{\mbox{cm}^{-2}}}
\newcommand{\ps}{\ensuremath{\mbox{s}^{-1}}}

\newcommand{\K}{\ensuremath{\mbox{K}}}

\newcommand{\asca}{\textit{ASCA}}
\newcommand{\chandra}{\textit{Chandra}}
\newcommand{\einstein}{\textit{Einstein}}
\newcommand{\rxte}{\textit{RXTE}}
\newcommand{\xmm}{\textit{XMM-Newton}}

\newcommand{\ArXVII}{Ar~\textsc{xvii}}
\newcommand{\ArXVIII}{Ar~\textsc{xviii}}
\newcommand{\CaXIX}{Ca~\textsc{xix}}
\newcommand{\CaXX}{Ca~\textsc{xx}}
\newcommand{\OVII}{O~\textsc{vii}}
\newcommand{\FeXXV}{Fe~\textsc{xxv}}
\newcommand{\FeXXVI}{Fe~\textsc{xxvi}}
\newcommand{\SXV}{S~\textsc{xv}}
\newcommand{\SXVI}{S~\textsc{xvi}}
\newcommand{\SiXIII}{Si~\textsc{xiii}}
\newcommand{\SiXIV}{Si~\textsc{xiv}}

\newcommand{\citepossessive}[1]{\citeauthor{#1}'s \citeyearpar{#1}}
\newcommand{\citetsq}[1]{\citeauthor{#1} [\citeyear{#1}]}

\newcommand{\ec}{$\eta$~Car}

\shorttitle{\textit{CHANDRA} GRATING SPECTROMETRY OF $\eta$ CARINAE}
\shortauthors{HENLEY ET AL.}

\begin{document}

\title{\textit{Chandra} X-ray Grating Spectrometry of $\eta$ Carinae near X-ray Minimum:
I. Variability of the Sulfur and Silicon Emission Lines}
\author{D.~B. Henley\altaffilmark{1},
	M.~F. Corcoran\altaffilmark{2,3},
	J.~M. Pittard\altaffilmark{4},
	I.~R. Stevens\altaffilmark{5},	
	K. Hamaguchi\altaffilmark{2,3}, and
	T.~R. Gull\altaffilmark{6}
}
\altaffiltext{1}{Department of Physics and Astronomy, University of Georgia, Athens, GA 30602; dbh@physast.uga.edu}
\altaffiltext{2}{NASA Goddard Space Flight Center, CRESST, Astrophysics Science Division, Code 662, Greenbelt, MD 20771}
\altaffiltext{3}{Universities Space Research Association, 10211 Wincopin Circle, Columbia, MD 21044}
\altaffiltext{4}{School of Physics and Astronomy, University of Leeds, Woodhouse Lane, Leeds, LS2 9JT, U.K.}
\altaffiltext{5}{School of Physics and Astronomy, University of Birmingham, Edgbaston, Birmingham, B15 2TT, U.K.}
\altaffiltext{6}{Astrophysics Science Division, Code 667, Goddard Space Flight Center, Greenbelt, MD 20771}

\begin{abstract}
We report on variations in
important X-ray emission lines in a series of \chandra\ grating spectra of the supermassive colliding wind binary star \ec, including key phases around the X-ray minimum/periastron passage in 2003.5.
The X-rays arise from the collision of the slow, dense wind of \ec\ with the fast, low-density wind of an otherwise hidden companion star. The X-ray emission lines provide
the only direct measure of the flow dynamics of the companion's wind along the wind-wind collision zone.  We concentrate here on the silicon and sulfur lines, which are the strongest and best resolved lines in the X-ray spectra.  
Most of the line profiles can be adequately fit with symmetric Gaussians with little significant skewness.  
Both the silicon and sulfur lines show significant velocity shifts and correlated increases in line widths through the observations.  
The $\R=$ forbidden-to-intercombination ratio from the \ion{Si}{13} and \ion{S}{15} triplets is near or above the low-density limit in all observations, suggesting that the line-forming region is $>1.6$ stellar radii from the companion star. 
We show that simple geometrical models cannot simultaneously fit both the observed centroid variations and changes in line width as a function of phase.  
We show that the observed profiles can be fitted with synthetic profiles with a reasonable
model of the emissivity along the wind-wind collision boundary. 
We use this analysis to help constrain the line formation region as a function of orbital phase, 
and the orbital geometry.
\end{abstract}


\section{INTRODUCTION}
\label{sec:Introduction}

The supermassive star \ec\ \citep{davidson97a} is notorious
for its extraordinarily large luminosity 
and its implicitly large mass \citep[$L>4\times10^{6}L_{\odot}$ and $M \sim 100 M_{\odot}$,][]{hillier01}, the beautiful bipolar ``Homunculus''
nebula which shrouds it \citep{gaviola50}, its wild
instability (most notably the ``Great Eruption'' of 1843 which created the Homunculus) and
its continued broad-band variations \citep{sterken96,davidson99b}.  Understanding \ec\ is important for a wide variety of
astrophysical topics regarding the formation and evolution of extremely massive stars, the processes by which such stars lose mass and angular momentum, and the ways in which they interact with their surroundings. 

\ec\ exhibits variability over a wide range of wavelengths,
from radio \citep{duncan03},
through infrared \citep{whitelock94,whitelock04,damineli96,damineli97,damineli00,davidson00},
optical \citep{steiner04}, 
and ultraviolet \citep{smith04a}
to X-rays \citep{ishibashi99,corcoran05b}.
All these variations have a characteristic cycle of almost exactly 2024 days, which strongly suggests that \ec\ is a 
long period ($P=2024$ day) binary 
\citep{damineli96,damineli97}.
The observed variability is believed to result (directly or indirectly) from the interaction of the fast wind
($\vc \sim 3000~\kmps$) and ionizing radiation from the companion with the dense, slow wind of the Luminous Blue Variable (LBV) primary
($\veta \sim 500~\kmps$). In this scheme, the X-rays are produced by the collision of the two stars' winds, which causes the
companion's fast wind to be shock-heated to tens of MK \citep{pittard98a,pittard02a}. The high temperature of the shocked wind of the companion explains the hard X-rays ($kT \ga 4~\kev$) first directly associated with the star
by \einstein\ \citep{seward79}.  Similar hard X-ray emission is seen from WR~140, the ``canonical'' long period
eccentric massive colliding wind binary \citep{pollock05}.

Our understanding of the system has become more sophisticated due in part to dense multiwavelength monitoring near the times of the X-ray minima in 1998 and 2003.5.  These observations showed that, at the same time that the X-ray brightness of the source reaches minimum, the ionization state of the circumstellar medium rapidly decreases \citep{1995ApJ...441L..73D,nielsen07}, the infrared \citep{whitelock04} and millimeter-wave \citep{2005A&A...437..977A} brightness of the source also drops, absorption components in excited \ion{He}{1} P-Cygni emission lines undergo rapid blue-to-red velocity shifts \citep{nielsen07}, \ion{He}{2} 4686-\angstrom\ emission \citep{steiner04,2006ApJ...640..474M} appears, shows a similar blue-to-red centroid shift, then disappears, and the far UV flux from \ec\ drops rapidly \citep{2005ApJ...633L..37I}. In X-rays, the hottest electron temperature stayed the same, but the ionization balance of Fe ions changed remarkably \citep{hamaguchi07}. In all colliding wind models these changes (which last only about 90 days of the 2024-day cycle) occur near periastron passage, and require a high eccentricity ($e\sim0.9$).  However, important details regarding the nature of the wind-wind collision are still not well constrained; there is still debate concerning, for example, whether the X-ray minimum occurs near inferior conjunction (when the companion is in front of the LBV primary) or superior conjunction; the magnitude of the companion's wind velocity; and the mass loss rates from either star. These uncertainties limit our understanding of how the companion star affects the system, and, ultimately limit our knowledge of the evolutionary state of the system.

The  detailed analysis of excited \ion{He}{1} P-Cygni absorption lines in
spatially resolved spectra by \citet{nielsen07}  showed
radial velocity variations which mimic the orbital radial velocity
variations expected in an eccentric ($e\approx0.9$) binary system
with the semi-major axis pointed towards the observer (longitude of
periastron $\omega\sim 270^{\circ}$) and an assumed inclination
$i\approx41^{\circ}$. These spectral variations suggest that the
ionized helium zone in the wind of the cool, massive primary star
approaches the observer prior to periastron passage.  They also
showed that the velocity amplitude of the \ion{He}{1} P-Cygni
absorption components ($\sim$140~\kmps) was larger than expected
if the absorption arises in the dense wind of the more massive star.
They concluded that the velocity variations are probably strongly
influenced by ionization effects due to the interaction of the
companion star's photospheric UV radiation with the wind of the cool
primary star.  They also suggested that some of the \ion{He}{1}
emission might originate within or near the wind-wind collision and
thus could be a diagnostic of that collision.  However the complex
influence of the companion's radiation with the primary wind makes
interpretation of such diagnostics far from straightforward.
(For an alternative explanation of the \ion{He}{1} observations, in which the lines are assumed to form in the acceleration zone of the secondary, see \citealp{kashi07}.)

X-ray line profiles provide the most direct probe of the dynamics of the wind of the unseen companion after it is shock-heated in the wind-wind interaction, since these lines originate in the high temperature plasma near the wind-wind shock interface. X-ray lines directly reflect the dynamic properties of this hot shocked gas.  In this paper we present our analysis of the high resolution X-ray grating spectra of \ec\ obtained by the High Energy Transmission Grating Spectrometer \citep[HETGS;][]{markert94} on the \chandra\ X-ray Observatory \citep{weisskopf02} obtained as part of a large observing campaign around the time of the 2003.5 X-ray minimum.  A preliminary  analysis of these data has appeared in \citet{henley05b}.  

In this paper
we discuss our analysis of  spectra in the energy range near 2~\kev\ obtained by the Medium and High Energy Gratings (MEG and HEG).   This energy range is dominated by line emission from Si and S hydrogen-like and helium-like ions.  These lines form in the cooler regions of the shocked gas farther along the wind-wind collision zone, and thus provide a better measure of the flow dynamics of the shock-heated wind of the companion along the colliding wind interface than the iron lines, which originate near the hottest part of the shock near the stagnation point where flow velocities are low. In this energy range the HETGS first order spectra has sufficient resolution to resolve the component lines of the He-like triplets providing useful density and temperature diagnostics. Unfortunately, potentially crucial line emission from C, N, and O (which could be used to measure abundances of the shocked companion's wind and help constrain the evolutionary state of the companion) are not observable in the central source due to the heavy absorption by the cold gas and dust in the Homunculus. 

This paper
is organized as follows.  The observations and the data reduction are described in \S\ref{sec:Observations}, and
the HETGS silicon and sulfur emission lines are discussed in \S\ref{sec:Spectra}. 
In \S\ref{sec:GeometricalModel}
we apply a simple geometrical model of the wind-wind collision to the variations 
in line centroids and widths.  
In \S\ref{sec:SyntheticProfiles} we apply synthetic colliding wind line profiles to the observed HETGS silicon and sulfur 
profiles.  We discuss the 
results of the emission line analysis in \S\ref{sec:Discussion}, 
and our conclusions are presented in \S\ref{sec:Conclusions}. 
Throughout this paper we quote $1\sigma$ errors.

\section{OBSERVATION DETAILS AND DATA REDUCTION}
\label{sec:Observations}

The details of the six \chandra\ HETGS observations of \ec\ are given in Table~\ref{tab:Observations}.
For the purposes of this paper we designate the observations with CXO, subscripted with the date in YYMMDD format
(cf.\ \citealp{hamaguchi07}). The earliest observation was in 2000 November
(CXO$_{001119}$; \citealp{corcoran01b,pittard02a}), approximately half-way between the
previous X-ray minimum in late 1997 and the X-ray minimum in mid-2003. The second observation was taken
approximately one year later (2002 October; CXO$_{021016}$), by which time the X-ray flux had increased
by a factor of $\sim$2. The four remaining observations were taken over the space of approximately five months
around the X-ray minimum which occurred, as expected, in late 2003 June. In particular, they approximately correspond
to X-ray maximum (2003 May; CXO$_{030502}$), the early part of the descent to X-ray minimum (2003 June; CXO$_{030616}$),
the X-ray minimum itself (2003 July; CXO$_{030720}$), and the recovery from the minimum (2003 September; CXO$_{030926}$).  All data were read out using the Advanced Camera for Imaging Spectroscopy spectroscopic array (ACIS-S).  The outer ACIS-S CCD chips (S0 and S5) were switched off, and we used a reduced read-out window in order to reduce pileup.  This truncates the low-energy spectra but results in little real data loss since the stellar source is heavily absorbed. 
The spectra at energies $E\lesssim3~\kev$ obtained during and just after the X-ray minimum (CXO$_{030720}$ and CXO$_{030926}$)
are contaminated by the ``Central Constant Emission" (CCE) component identified by \citet{hamaguchi07} from \xmm\
observations taken during the 2003 X-ray minimum. This means that the silicon and sulfur lines from these two spectra do not accurately
reflect the emission from the colliding wind plasma alone (with the exception of \SXVI\ in CXO$_{030926}$, which is not as badly contaminated).
However, for completeness, we include measurements of the line properties
for all six observations, including CXO$_{030720}$ and CXO$_{030926}$, in our  discussion in \S\S\ref{sec:Spectra} and \ref{sec:GeometricalModel}.

\begin{deluxetable*}{cccccccccc}
\tabletypesize{\scriptsize}
\tablewidth{0pt}
\tablecaption{\chandra-HETGS Observations of $\eta$ Carinae\label{tab:Observations}}
\tablehead{
\colhead{Observation}			& \colhead{Observation}			& \colhead{Start}	& \colhead{Phase\tablenotemark{c}}	& \colhead{Exposure}	& \multicolumn{2}{c}{HEG}						& \multicolumn{2}{c}{MEG} \\
\colhead{ID\tablenotemark{a}}		& \colhead{ID\tablenotemark{b}}		& \colhead{date}	& \colhead{$\phi$}			& \colhead{(ks)}	& \colhead{Counts\tablenotemark{d}}	& \colhead{Rate (s$^{-1}$)}	& \colhead{Counts\tablenotemark{d}}	& \colhead{Rate (s$^{-1}$)}
}
\startdata
CXO$_{001119}$				& 632					& 2000 Nov 19		& 0.528					& 89.5			& 18459					& 0.206				& 20772					& 0.232 \\
CXO$_{021016}$				& 3749					& 2002 Oct 16		& 0.872					& 91.2			& 38160					& 0.418				& 45038					& 0.493 \\
CXO$_{030502}$				& 3745					& 2003 May 2		& 0.970					& 94.5			& 78264					& 0.828				& 81925					& 0.867 \\
CXO$_{030616}$				& 3748					& 2003 Jun 16		& 0.992					& 97.2			& 42411					& 0.436				& 40553					& 0.417 \\
CXO$_{030720}$				& 3746					& 2003 Jul 20		& 1.009					& 90.3			& 1183					& 0.013				& 1725					& 0.019 \\
CXO$_{030926}$				& 3747					& 2003 Sep 26		& 1.043					& 70.1			& 11137					& 0.159				& 8098					& 0.116 \\
\enddata
\tablenotetext{a}{Observation identification used in this paper (after \citealp{hamaguchi07}).\\ }
\tablenotetext{b}{Official \chandra\ observation identification.\\ }
\tablenotetext{c}{Mid-observation phase, calculated using the emphemeris in \citet{corcoran05b}.\\ }
\tablenotetext{d}{Total number of first-order ($+1$ and $-1$) non-background-subtracted counts.\\ }
\end{deluxetable*}

The data for all six observations were reduced from the Level 1 events files using CIAO\footnote{http://cxc.harvard.edu/ciao}  v3.4 and CALDB v3.3.0.1.  These versions 
are much improved over the earlier versions used by \citet{corcoran01b} and \citet{henley05b}.  We
followed the threads available from the \chandra\ website\footnote{http://cxc.harvard.edu/ciao/threads}.
We first removed the 
\texttt{acis\_detect\_afterglow} correction, and generated a new bad pixel file using \texttt{acis\_run\_hotpix}.
We then reprocessed the Level 1 events file with the latest
calibration using \texttt{acis\_process\_events}. This applies a new ACIS gain map, the time-dependent ACIS gain correction,
the ACIS charge transfer inefficiency (CTI) correction, and pixel and PHA randomization.
We then used \texttt{tgdetect} to determine the position of the zeroth-order image of \ec, \texttt{tg\_create\_mask} to determine
the location of the HEG and MEG ``arms'', and \texttt{tg\_resolve\_events} to assign the measured events to the different spectral orders.
After applying grade filters (\asca\ grades 0, 2, 3, 4, and 6 were kept) and good time intervals, we used \texttt{destreak} to remove
streaks           caused by a flaw in the serial readout which randomly deposits significant amounts of charge along the pixel row as charge is read out. Finally, we used \texttt{tgextract} to extract the grating spectra from the events file.
Spectral response files were also generated following the \chandra\ threads: we generated redistribution matrix files (RMFs)
and ancillary response files (ARFs) using \texttt{mkgrmf} and \texttt{fullgarf}, respectively.

\begin{figure}
\centering
\includegraphics[width=0.9\linewidth]{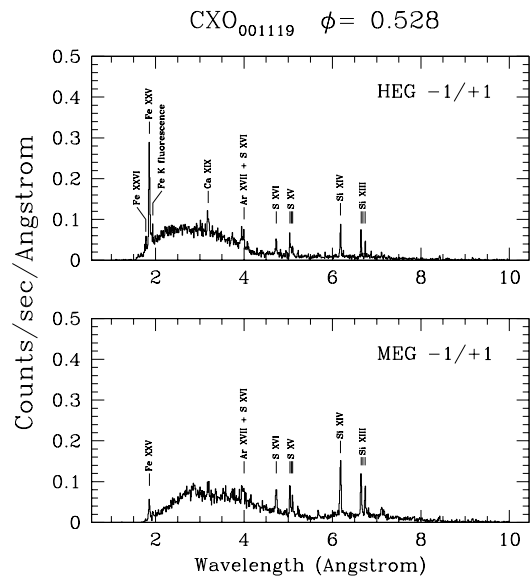}
\caption{\chandra\ HETGS spectra of \ec\ from CXO$_{001119}$. For each grating (HEG and MEG) the $+1$ and $-1$ orders
have been co-added, and the spectra have been binned up to 0.01~\angstrom.\label{fig:632spectrum}}
\end{figure}

\begin{figure}
\centering
\includegraphics[width=0.9\linewidth]{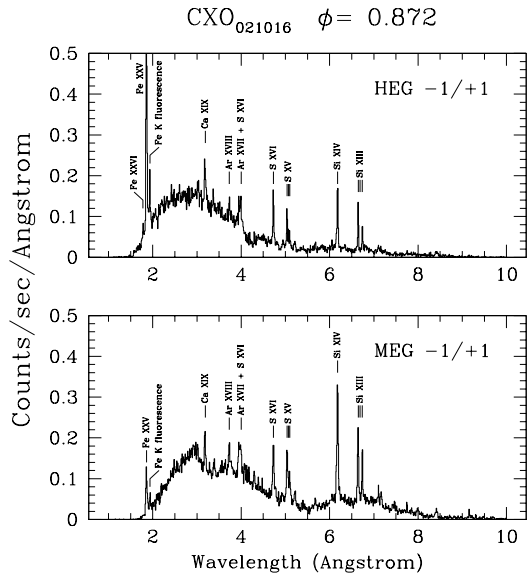}
\caption{As Figure~\ref{fig:632spectrum}, but for CXO$_{021016}$.\label{fig:3749spectrum}}
\end{figure}

\begin{figure}
\centering
\includegraphics[width=0.9\linewidth]{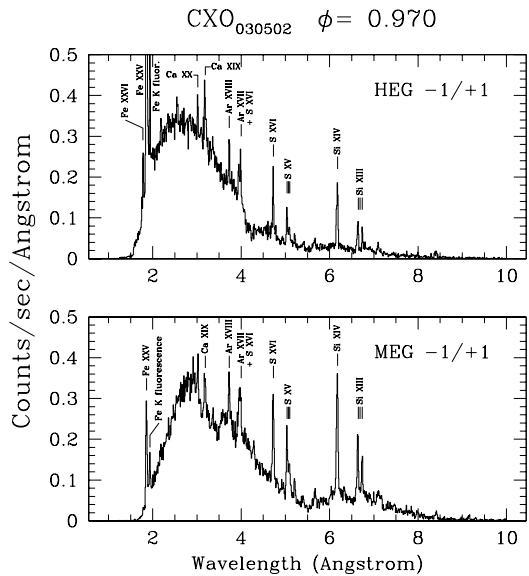}
\caption{As Figure~\ref{fig:632spectrum}, but for CXO$_{030502}$.\label{fig:3745spectrum}}
\end{figure}

\begin{figure}
\centering
\includegraphics[width=0.9\linewidth]{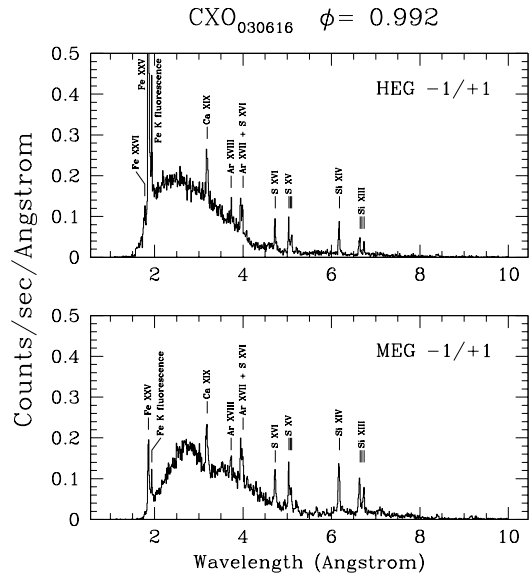}
\caption{As Figure~\ref{fig:632spectrum}, but for CXO$_{030616}$.\label{fig:3748spectrum}}
\end{figure}

\begin{figure}
\centering
\includegraphics[width=0.9\linewidth]{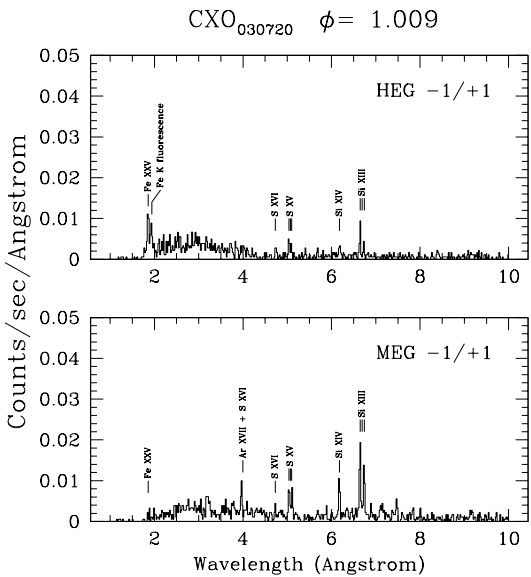}
\caption{As Figure~\ref{fig:632spectrum}, but for CXO$_{030720}$, with the spectrum binned 
up to 0.02~\angstrom, instead of 0.01~\angstrom. Note that the $y$-axis range is different from
Figures~\ref{fig:632spectrum} through \ref{fig:3748spectrum} and \ref{fig:3747spectrum}.
\label{fig:3746spectrum}}
\end{figure}

\begin{figure}
\centering
\includegraphics[width=0.9\linewidth]{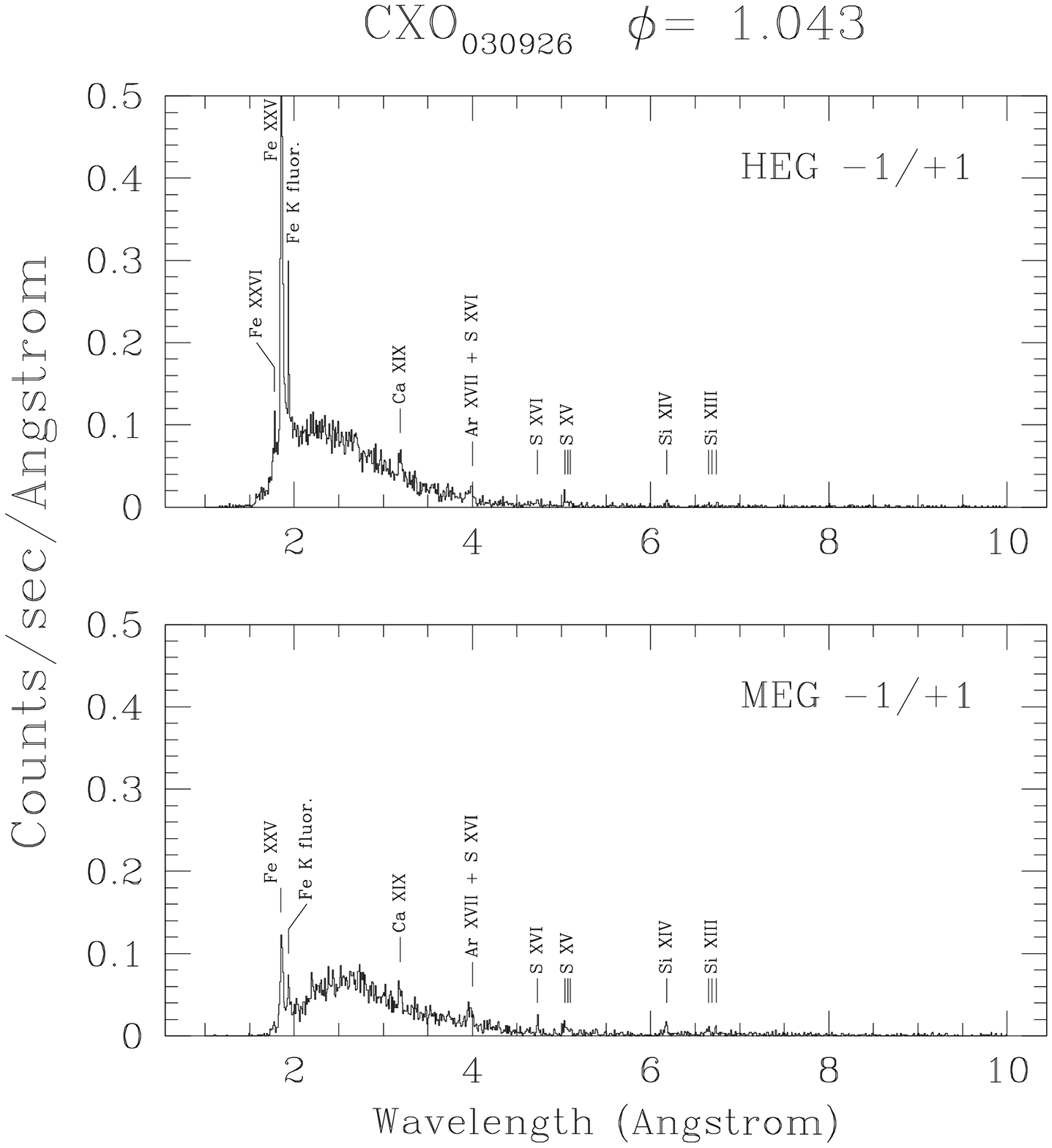}
\caption{As Figure~\ref{fig:632spectrum}, but for CXO$_{030926}$.\label{fig:3747spectrum}}
\end{figure}

The \chandra\ HETGS spectra of \ec\ for each of the six \chandra\ observations are shown in Figures~\ref{fig:632spectrum}
through \ref{fig:3747spectrum}. For each observation,
the $+1$ and $-1$ orders of each grating (HEG and MEG) have been co-added, and the spectra have been binned up to 0.01~\angstrom, except
for the observation taking during the X-ray minimum (CXO$_{030720}$; Fig.~\ref{fig:3746spectrum}), which has been binned up to 0.02~\angstrom.
Note that the spectra are shown with the same $y$-axis range, except for CXO$_{030720}$.

With the exception of CXO$_{030720}$, which is the faintest spectrum by an order of magnitude, the spectra all exhibit prominent continuum
emission and numerous emission lines. Particularly prominent are forbidden-intercombination-resonance (f-i-r) triplets from He-like
\FeXXV\  ($\lambda_\mathrm{resonance} = 1.85$~\angstrom), \SXV\ ($\lambda_\mathrm{res} = 5.04$~\angstrom) 
and \SiXIII\ ($\lambda_\mathrm{res} = 6.65$~\angstrom),
\Lyalpha\ emission from H-like \SXVI\ ($\lambda = 4.73$~\angstrom) and \SiXIV\ ($\lambda = 6.18$~\angstrom),
and K-shell fluorescent emission from cool Fe ($\lambda = 1.94$~\angstrom).
Other lines which are visible (not necessarily in all spectra) include \CaXX\ at 3.0~\angstrom, \CaXIX\ at 3.2~\angstrom, \ArXVIII\
at 3.7~\angstrom, \ArXVII\ $+$ \SXVI\ at 4.0~\angstrom, \SiXIV\ at 5.2~\angstrom, and \SiXIII\ at 5.7~\angstrom.
The Fe K lines will be discussed elsewhere (Paper II; M.~F. Corcoran et~al., in preparation). Here we concentrate on the brightest
of the lower-excitation lines: the H-like \Lyalpha\ line and He-like f-i-r triplet from Si and S. Although line shifts and widths
can be measured for some of the other lines, the four lines that we concentrate on here are the only ones for which results can be obtained
from all six observations. Furthermore, the analysis of these weaker lines is consistent with the analysis of the stronger lines presented here 
\citep[for more detailed discussion of these weaker lines see][]{henley05b}.

\section{SILICON AND SULFUR LINE PROFILES}
\label{sec:Spectra}

\subsection{Gaussian Modeling}
\label{subsec:Gaussian}

We analyzed the \chandra\ spectra of \ec\ using unbinned,
non-co-added spectra, so no spectral information was lost. Because some bins contain low numbers of counts,
the Cash statistic \citep{cash79} was used instead of the $\chi^2$ statistic.
To measure each emission line's properties, we analyzed each line (or multiplet) individually over a narrow range
of wavelengths encompassing just the line of interest. We then fit a model to the data comprising a power-law continuum
component plus Gaussian components to model the line emission. The number of Gaussians used, and how their
parameters are tied together, depended on the nature of the line being analyzed \citep{pollock05,henley05a}.
For the \Lyalpha\ lines (which are closely spaced doublets, separated by $\approx$5~\mA), we used two Gaussians.
The Doppler shifts of the two components were constrained to be equal, as were their widths, and the intensity
of the longer-wavelength component was constrained to be half that of the shorter-wavelength component. The He-like
f-i-r triplets were fit with three Gaussians, the Doppler shifts and widths of which were tied together as for the
Lyman lines, but the amplitudes of which were allowed to vary. For the intercombination line we used the rest wavelength
of the $2\,{^3\mathrm{P}_1} \rightarrow 1\,{^1\mathrm{S}_0}$ transition, and ignored the fainter
$2\,{^3\mathrm{P}_2} \rightarrow 1\,{^1\mathrm{S}_0}$ transition.

The analysis described here was carried out using SHERPA,
as distributed with CIAO v3.4. The data were not background subtracted, as the Cash statistic cannot
be used on background-subtracted data, nor was the background separately modeled out. This is not a problem
because for the lines of interest the background count rate is 
more than an order of magnitude lower than the source count rate in the relevant energy range. Furthermore,
the background spectra show no prominent spectral features, so any background contribution would be included in
the continuum component used in the fitting.

Our procedure for a given line from a given observation was to fit the same model to all four spectra
(HEG~$\pm 1$, MEG~$\pm 1$) simultaneously. We then assessed goodness-of-fit using a Monte Carlo method (as the
Cash statistic by itself gives no goodness of fit information), using a similar method to that of \citet{helsdon00}.
The best-fit model was used to simulate 1000 synthetic emission lines. Poisson noise was added to each simulated line,
and then each was compared with the original model to calculate its Cash statistic. Hence, for a given emission line model,
we obtained the distribution of Cash statistic values expected for datasets generated from that model. By comparing the
observed Cash statistic with this distribution, we determined the probability that the model could have produced
the observed data. In practice we did this by measuring the mean and standard deviation ($\sigma$) of the simulated Cash statistic values --
if the observed Cash statistic lay more than $2\sigma$ away from the mean, the fit was regarded as ``poor''.

We found that, when fitting to all four spectra simultaneously, Gaussian profiles gave acceptable fits to most of the lines. 
A visual inspection of the poorer fits indicated that the lines in different orders were sometimes slightly offset from each other
in wavelength. This may be due to uncertainty in the determination of the centroid position of the
zeroth-order image on the ACIS-S detector -- if the determined position were offset from the true position, the wavelengths
in the $+1$ and $-1$ orders would be offset in opposite directions. To overcome this, where possible we fit the model
to the four spectra individually, and then averaged the results. For some fainter lines (the sulfur lines in CXO$_{001119}$, and
the lines in CXO$_{030720}$ and CXO$_{030926}$) we were unable to constrain the model in all four individual spectra. In these cases,
we adopted
the results obtained by fitting all four spectra simultaneously. For the \SiXIII\ triplet in
CXO$_{030720}$, even this did not work, and instead we obtained our results by fitting just to the MEG~$+1$ and $-1$ spectra.

\begin{figure*}
\centering
\includegraphics[width=0.4\linewidth]{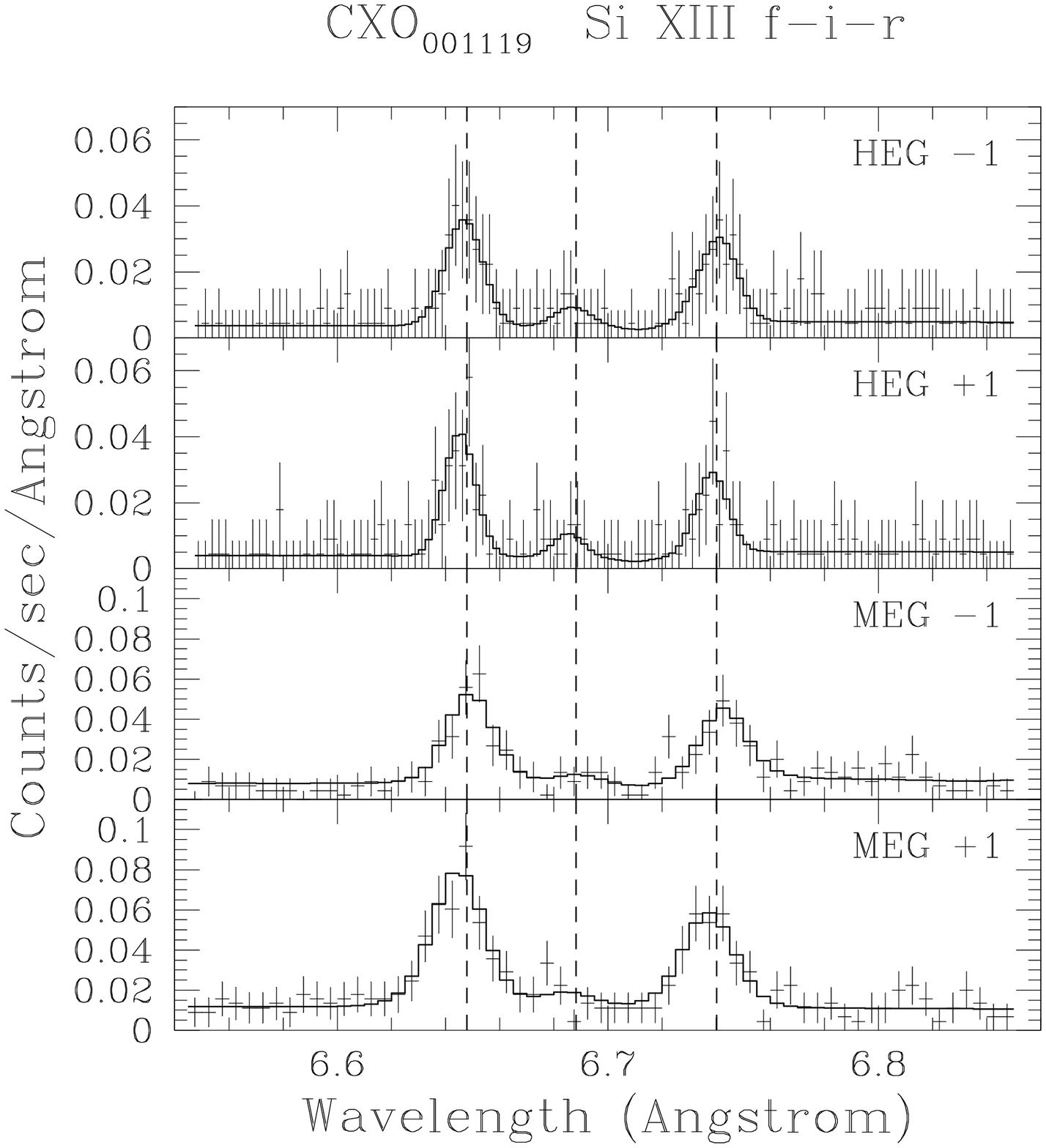}
\hspace{0.1\linewidth}
\includegraphics[width=0.4\linewidth]{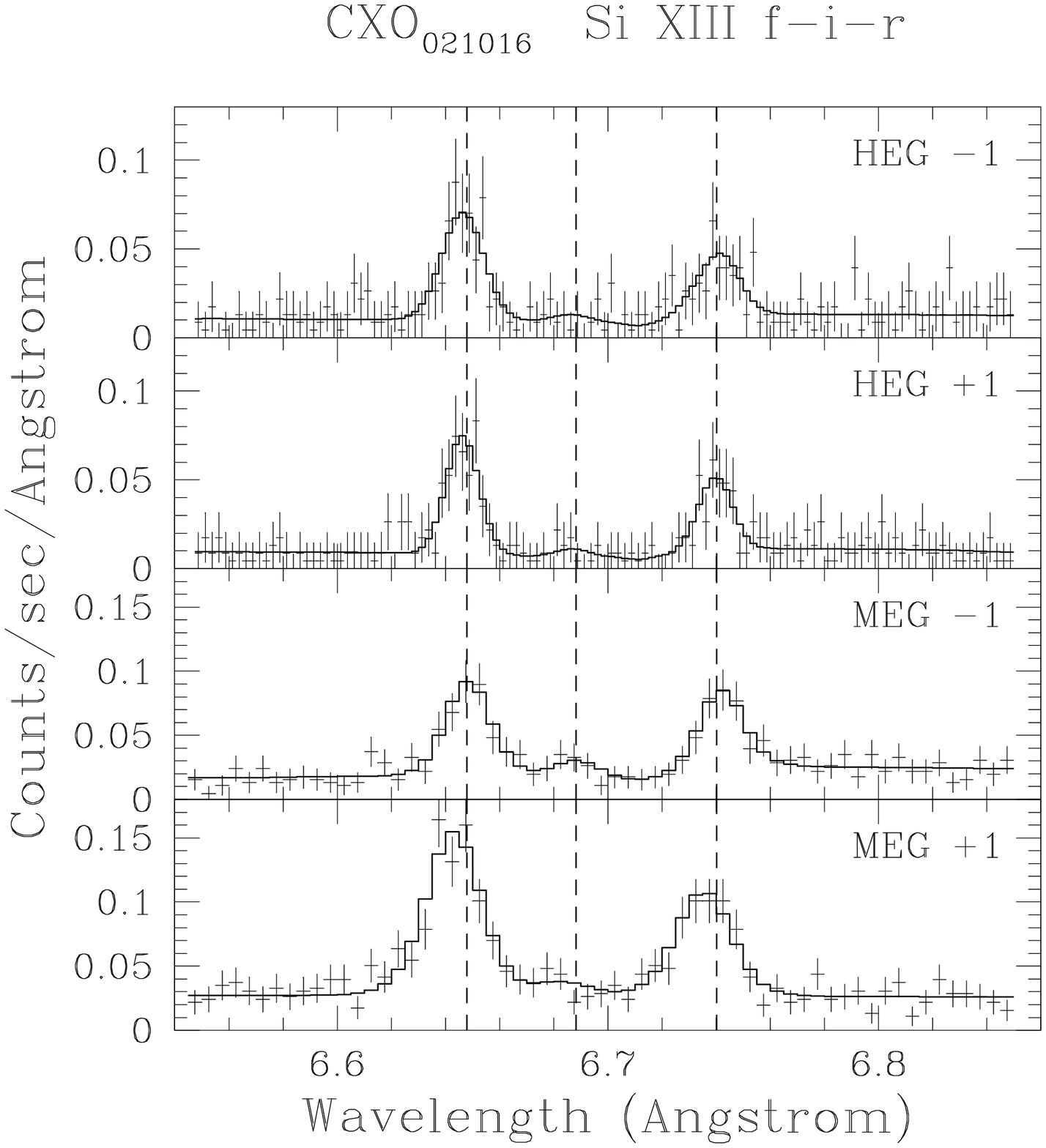} \\
\vspace{1cm}
\includegraphics[width=0.4\linewidth]{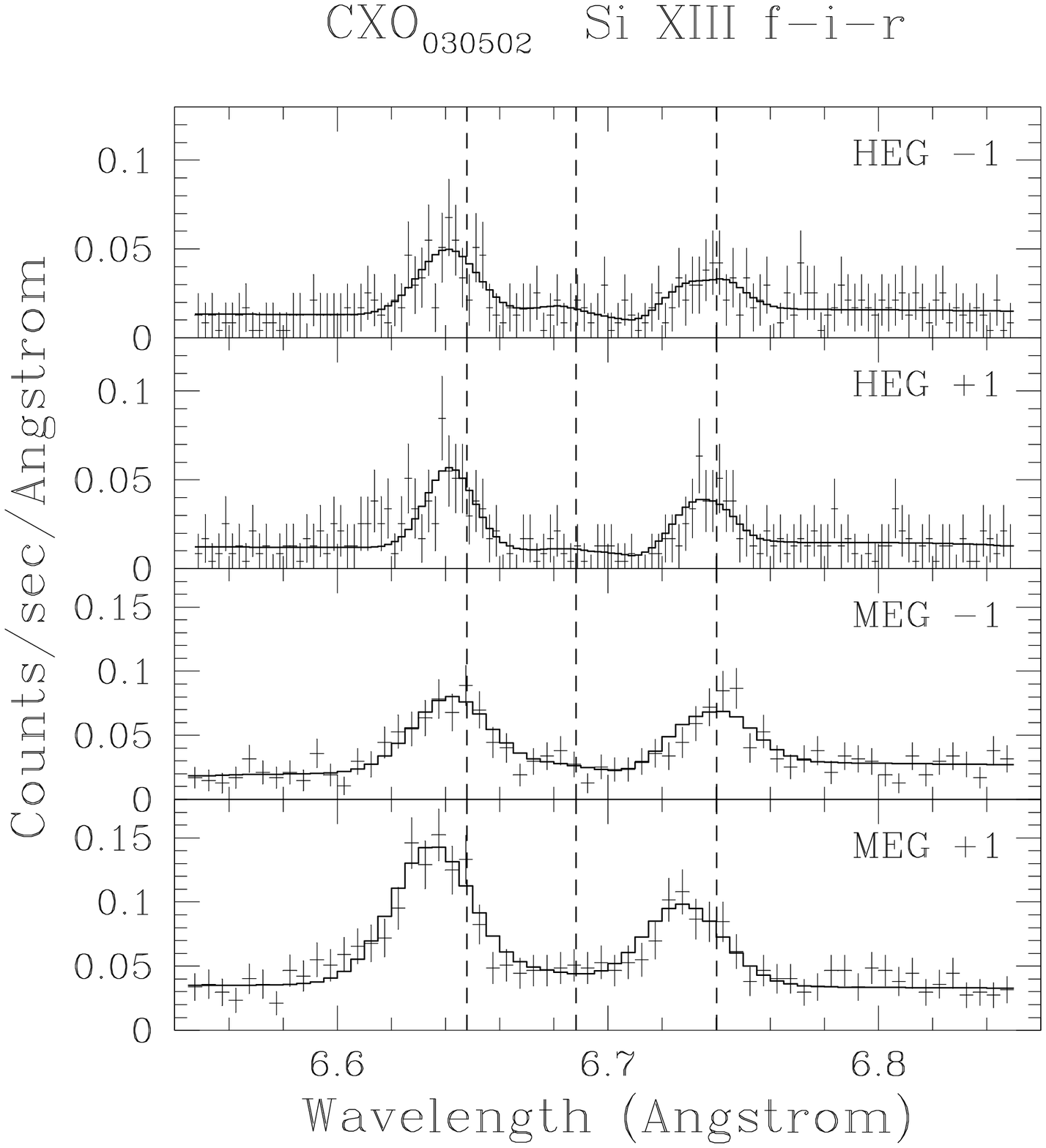}
\hspace{0.1\linewidth}
\includegraphics[width=0.4\linewidth]{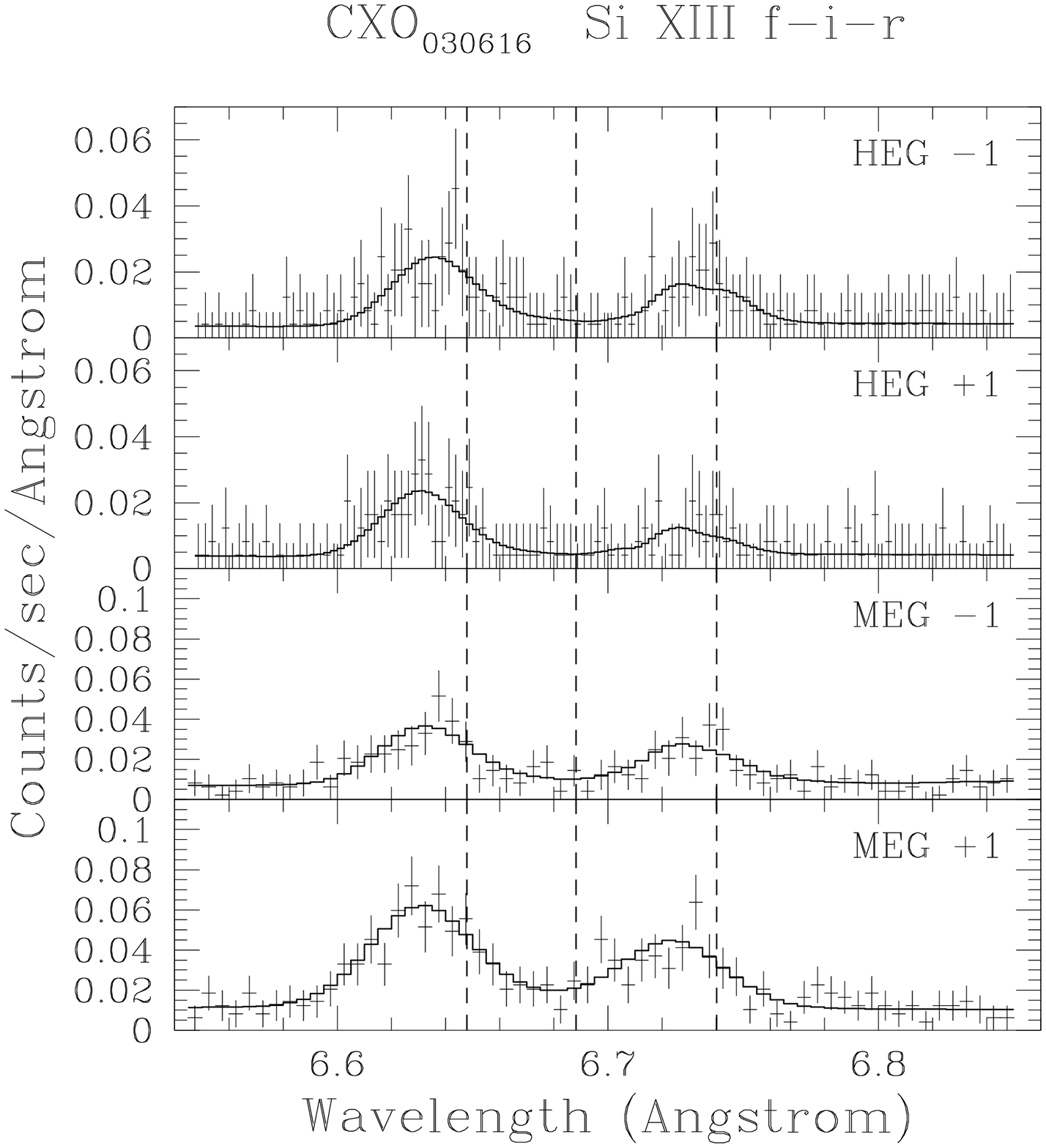}
\caption{\SiXIII\ f-i-r triplets from CXO$_{001119}$, CXO$_{021016}$, CXO$_{030502}$, and CXO$_{030616}$, showing the best-fitting Gaussian line model obtained
by fitting to each spectral order individually. The vertical dashed lines show the rest wavelengths of the resonance, intercombination, and forbidden lines
(6.6479, 6.6882, and 6.7403~\angstrom, respectively).
\label{fig:SiXIII}}
\end{figure*}

\begin{figure*}
\centering
\includegraphics[width=0.4\linewidth]{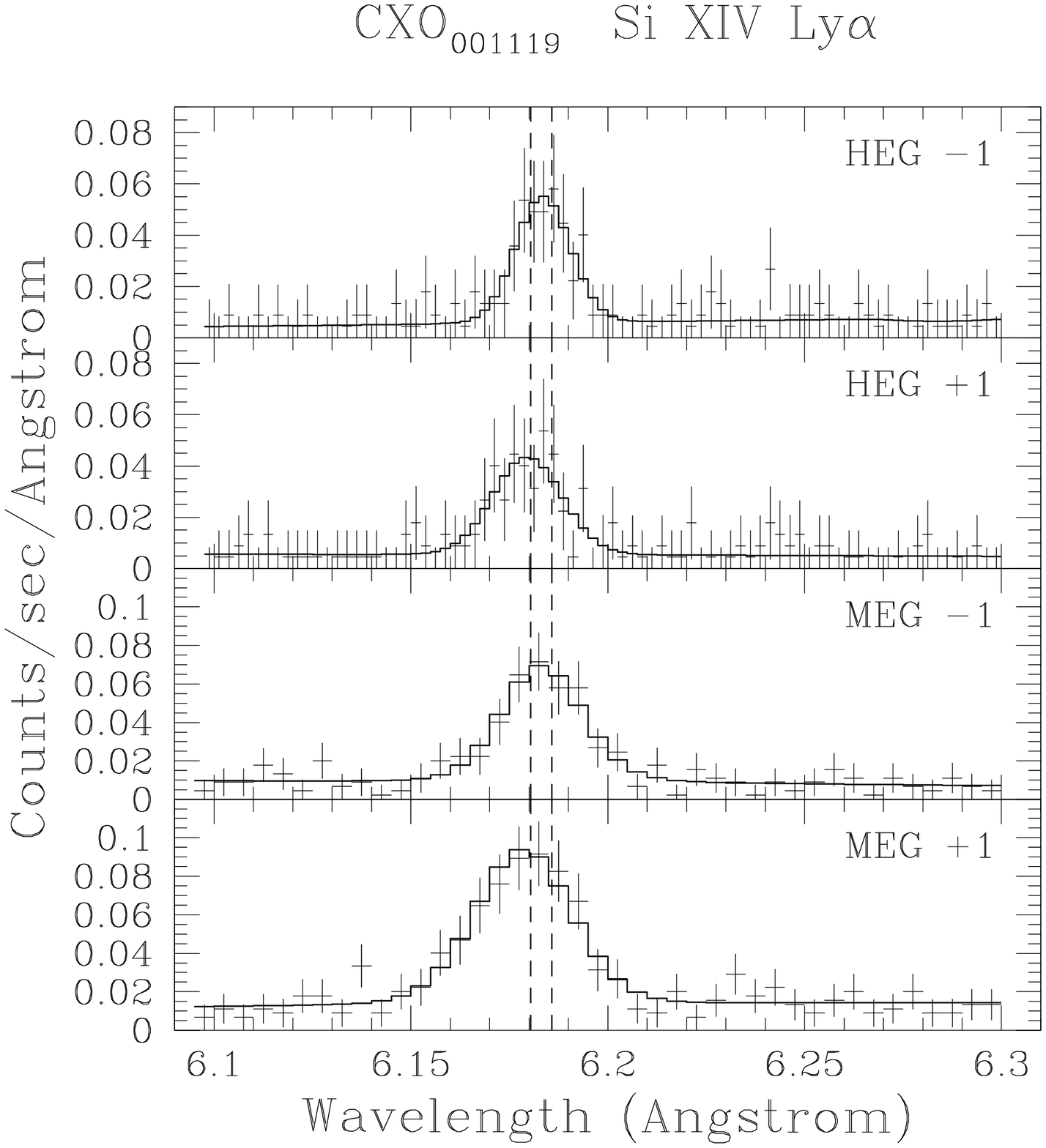}
\hspace{0.1\linewidth}
\includegraphics[width=0.4\linewidth]{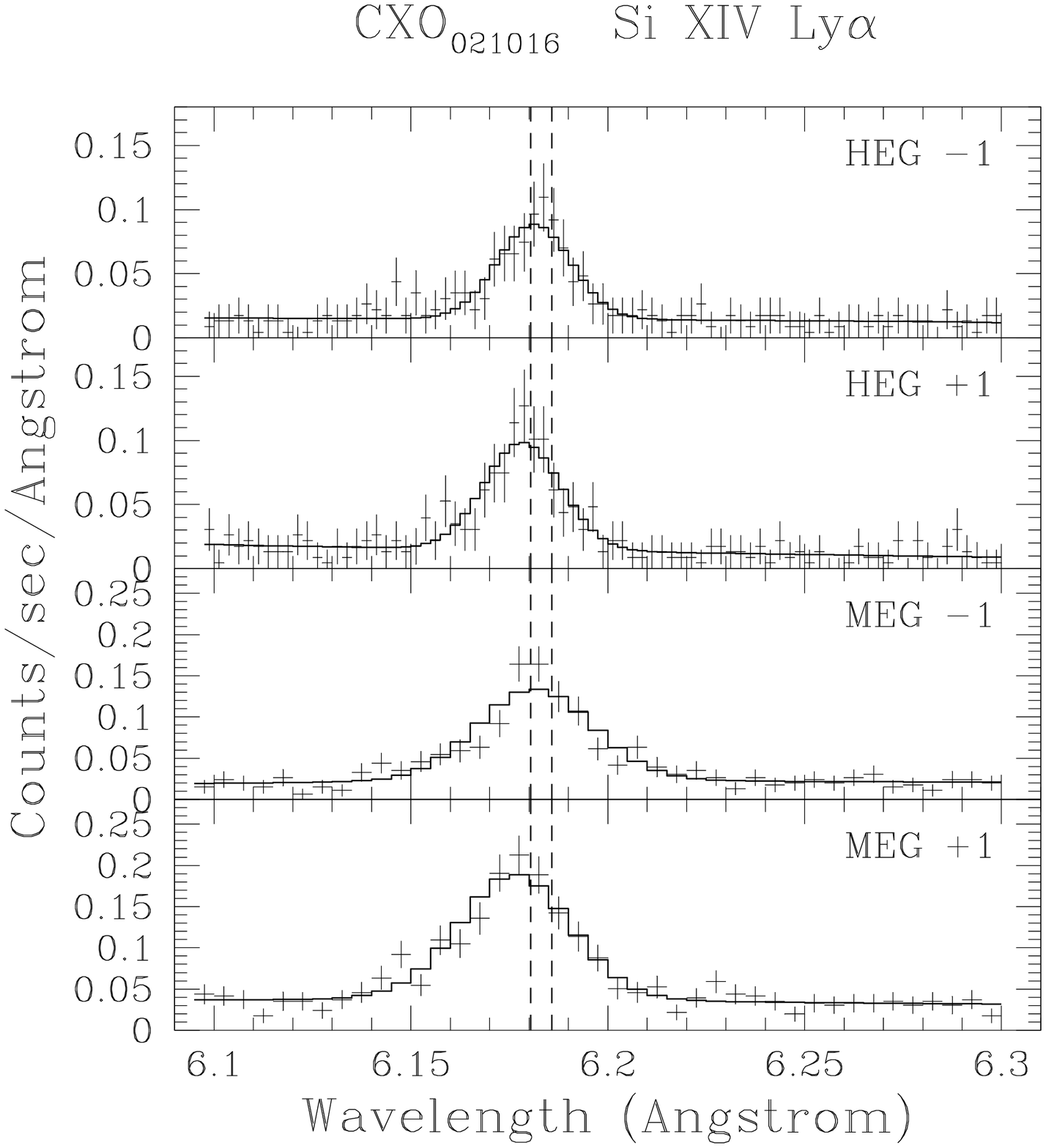} \\
\vspace{1cm}
\includegraphics[width=0.4\linewidth]{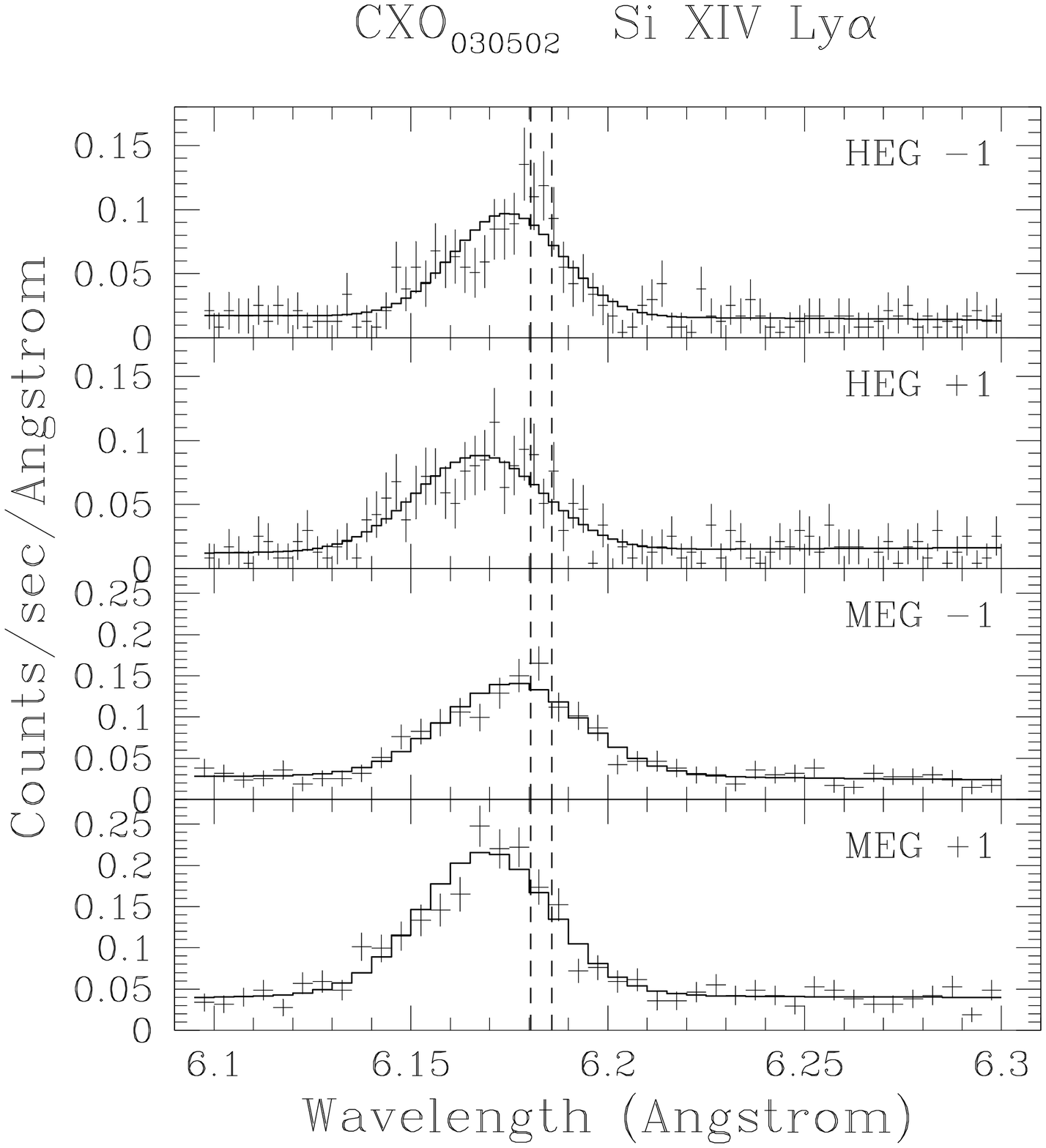}
\hspace{0.1\linewidth}
\includegraphics[width=0.4\linewidth]{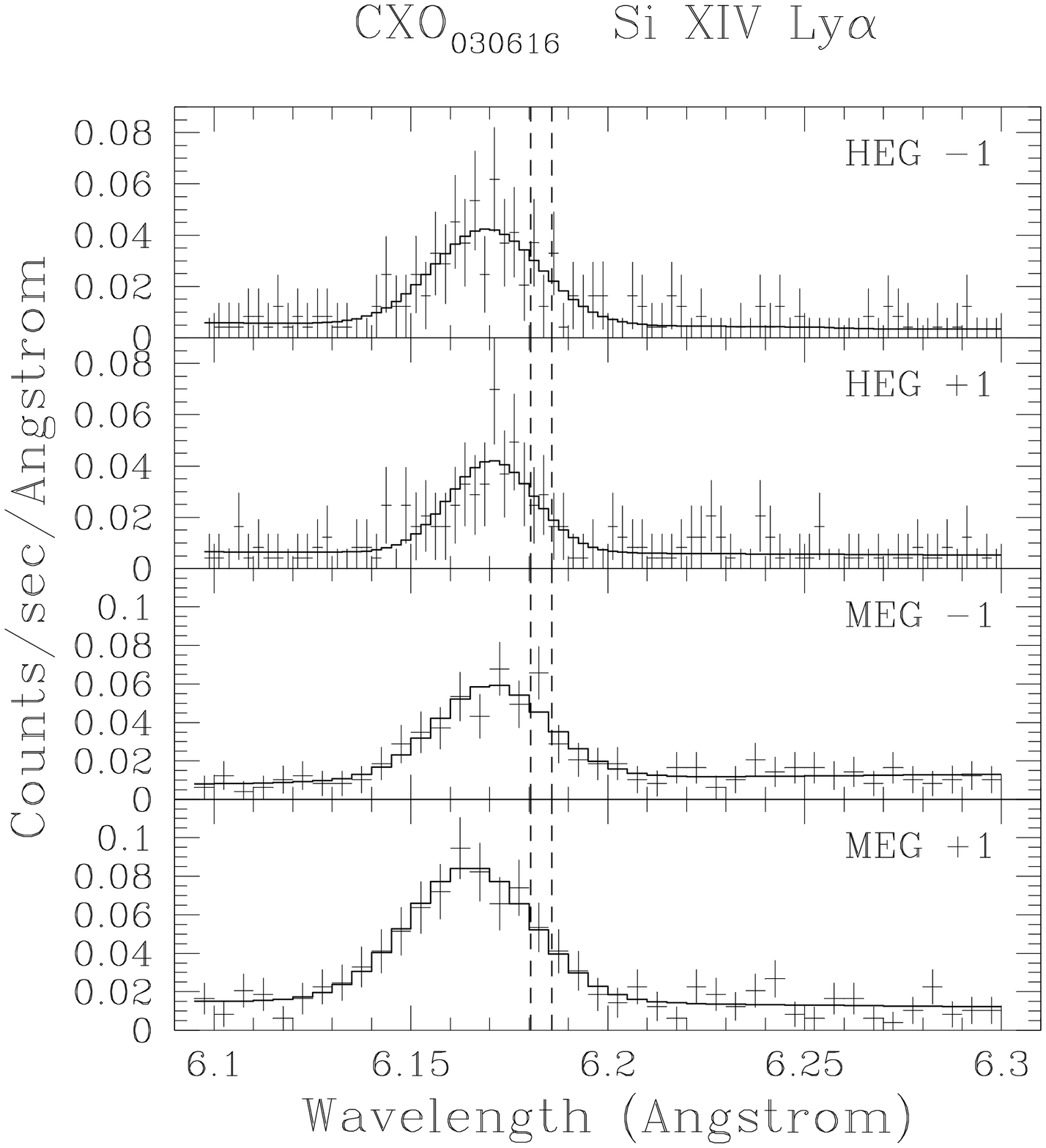}
\caption{As Figure~\ref{fig:SiXIII}, but showing the \SiXIV\ \Lyalpha\ line. The vertical dashed lines show the rest wavelengths of the two components of the
line (6.1804 and 6.1858~\angstrom, respectively).
\label{fig:SiXIV}}
\end{figure*}

The emission line shifts, widths, fluxes, and equivalent widths measured from this Gaussian modeling are given in Table~\ref{tab:LineResults}.
The rest wavelengths are adopted
from ATOMDB\footnote{http://cxc.harvard.edu/atomdb} v1.3.1.
Table~\ref{tab:LineVelocities} shows the results in Table~\ref{tab:LineResults} expressed as velocities.
Figures~\ref{fig:SiXIII} and \ref{fig:SiXIV} show the \SiXIII\ and \SiXIV\ lines from the four brightest spectra
(CXO$_{001119}$, CXO$_{021016}$, CXO$_{030502}$, and CXO$_{030616}$), along with the best-fitting Gaussian line model. The models were fit to each
spectral order individually, which is why in several cases the Gaussians are offset in the different orders.

\begin{deluxetable*}{llc r@{\,$\pm$\,}l r@{\,$\pm$\,}l r@{\,$\pm$\,}l r@{\,$\pm$\,}l c}
\tabletypesize{\scriptsize}
\tablewidth{0pt}
\tablecaption{Emission Line Fit Results\label{tab:LineResults}}
\tablehead{
\colhead{Ion}	& \colhead{Line}	& \colhead{$\lambda_0$}			& \multicolumn{2}{c}{$\lambda_\mathrm{obs}$}	& \multicolumn{2}{c}{$\Delta \lambda$ (FWHM)}	& \multicolumn{2}{c}{Flux} 				& \multicolumn{2}{c}{EW}	& \colhead{Fitting} \\
		&			& \colhead{(\angstrom)}			& \multicolumn{2}{c}{(\angstrom)}		& \multicolumn{2}{c}{(\mA)}			& \multicolumn{2}{c}{($10^{-5}$ ph \pcmsq\ \ps)}	& \multicolumn{2}{c}{(\AA)}	& \colhead{method}  \\
\colhead{(1)}	& \colhead{(2)}		& \colhead{(3)}				& \multicolumn{2}{c}{(4)}			& \multicolumn{2}{c}{(5)}			& \multicolumn{2}{c}{(6)}				& \multicolumn{2}{c}{(7)}	& \colhead{(8)}
}
\startdata
\multicolumn{10}{c}{CXO$_{001119}$} \\
\hline
\SXVI		& \Lyalpha		& 4.7274 				& 4.7267 & 0.0012				& \phn20.2 & 4.9\phn				& \phantom{00000}2.96 & 0.34\phantom{00000}		& 0.041 & 0.005			& (a) \\
\SXV		& r			& 5.0387				& 5.0366 & 0.0007				& 13.7 & 1.8					& 4.99 & 0.41						& 0.087 & 0.007			& (a) \\
		& i			& 5.0665				& \multicolumn{2}{c}{5.0644}			& \multicolumn{2}{c}{13.8}			& 0.97 & 0.27						& 0.017 & 0.005			& (a) \\
		& f			& 5.1015				& \multicolumn{2}{c}{5.0994}			& \multicolumn{2}{c}{13.9}			& 2.41 & 0.33						& 0.042 & 0.006			& (a) \\
\SiXIV		& \Lyalpha		& 6.1804				& 6.1797 & 0.0006				& 16.5 & 1.7					& 2.43 & 0.13						& 0.117 & 0.006			& (b) \\
\SiXIII		& r			& 6.6479				& 6.6461 & 0.0005				& 12.3 & 1.3					& 2.57 & 0.16						& 0.171 & 0.011			& (b) \\
		& i			& 6.6882				& \multicolumn{2}{c}{6.6864}			& \multicolumn{2}{c}{12.4}			& 0.35 & 0.09						& 0.024 & 0.006			& (b) \\
		& f			& 6.7403				& \multicolumn{2}{c}{6.7385}			& \multicolumn{2}{c}{12.5}			& 1.75 & 0.13						& 0.121 & 0.009			& (b) \\
\hline
\multicolumn{10}{c}{CXO$_{021016}$} \\
\hline
\SXVI		& \Lyalpha		& 4.7274 				& 4.7262 & 0.0006				& 14.1 & 2.0					& 6.97 & 0.44						& 0.042 & 0.003			& (b) \\
\SXV		& r			& 5.0387				& 5.0374 & 0.0006				& 13.6 & 1.5					& 9.27 & 0.67						& 0.068 & 0.005			& (b) \\
		& i			& 5.0665				& \multicolumn{2}{c}{5.0652}			& \multicolumn{2}{c}{13.7}			& 3.08 & 0.49						& 0.022 & 0.004			& (b) \\
		& f			& 5.1015				& \multicolumn{2}{c}{5.1002}			& \multicolumn{2}{c}{13.8}			& 3.79 & 0.53						& 0.028 & 0.004			& (b) \\
\SiXIV		& \Lyalpha		& 6.1804				& 6.1772 & 0.0005				& 23.3 & 1.6					& 5.62 & 0.23						& 0.103 & 0.004			& (b) \\
\SiXIII		& r			& 6.6479				& 6.6458 & 0.0004				& 13.8 & 1.1					& 4.89 & 0.23						& 0.133 & 0.006			& (b) \\
		& i			& 6.6882				& \multicolumn{2}{c}{6.6861}			& \multicolumn{2}{c}{13.9}			& 0.50 & 0.12						& 0.014 & 0.003			& (b) \\
		& f			& 6.7403				& \multicolumn{2}{c}{6.7382}			& \multicolumn{2}{c}{14.0}			& 3.08 & 0.18						& 0.087 & 0.005			& (b) \\
\hline
\multicolumn{10}{c}{CXO$_{030502}$} \\
\hline
\SXVI		& \Lyalpha		& 4.7274				& 4.7221 & 0.0006				& 21.0 & 1.9					& 11.38 & 0.58						& 0.043 & 0.002			& (b) \\
\SXV		& r			& 5.0387				& 5.0335 & 0.0009				& 22.1 & 2.0					& 11.45 & 0.78						& 0.066 & 0.005			& (b) \\
		& i			& 5.0665				& \multicolumn{2}{c}{5.0613}			& \multicolumn{2}{c}{22.2}			& 3.34  & 0.59						& 0.017 & 0.004			& (b) \\
		& f			& 5.1015				& \multicolumn{2}{c}{5.0962}			& \multicolumn{2}{c}{22.4}			& 5.71 & 0.68						& 0.027 & 0.004			& (b) \\
\SiXIV		& \Lyalpha		& 6.1804				& 6.1698 & 0.0005				& 33.7 & 1.4					& 7.68 & 0.26						& 0.125 & 0.004			& (b) \\
\SiXIII		& r			& 6.6479				& 6.6397 & 0.0006				& 24.6 & 1.6					& 5.55 & 0.27						& 0.122 & 0.005			& (b) \\
		& i			& 6.6882				& \multicolumn{2}{c}{6.6800}			& \multicolumn{2}{c}{24.7}			& 0.62 & 0.16						& 0.014 & 0.004			& (b) \\
		& f			& 6.7403				& \multicolumn{2}{c}{6.7322}			& \multicolumn{2}{c}{24.9}			& 3.52 & 0.22						& 0.080 & 0.005			& (b) \\
\hline
\multicolumn{10}{c}{CXO$_{030616}$} \\
\hline
\SXVI		& \Lyalpha		& 4.7274				& 4.7213 & 0.0009				& 22.6 & 2.8					& 4.96 & 0.39						& 0.051 & 0.004			& (b) \\
\SXV		& r			& 5.0387				& 5.0312 & 0.0008				& 21.5 & 2.2					& 8.18 & 0.56						& 0.113 & 0.008			& (b) \\
		& i			& 5.0665				& \multicolumn{2}{c}{5.0590}			& \multicolumn{2}{c}{21.6}			& 2.50 & 0.41						& 0.036 & 0.006			& (b) \\
		& f			& 5.1015				& \multicolumn{2}{c}{5.0939}			& \multicolumn{2}{c}{21.8}			& 4.63 & 0.45						& 0.068 & 0.007			& (b) \\
\SiXIV		& \Lyalpha		& 6.1804				& 6.1667 & 0.0007				& 28.3 & 1.9					& 2.92 & 0.15						& 0.130 & 0.007			& (b) \\
\SiXIII		& r			& 6.6479				& 6.6329 & 0.0009				& 35.7 & 2.2					& 3.42 & 0.20						& 0.244 & 0.014			& (b) \\
		& i			& 6.6882				& \multicolumn{2}{c}{6.6731}			& \multicolumn{2}{c}{35.9}			& 0.34 & 0.13						& 0.025 & 0.010			& (b) \\
		& f			& 6.7403				& \multicolumn{2}{c}{6.7251}			& \multicolumn{2}{c}{36.2}			& 2.30 & 0.16						& 0.172 & 0.012			& (b) \\
\hline
\multicolumn{10}{c}{CXO$_{030720}$} \\
\hline
\SXVI		& \Lyalpha		& 4.7274				& 4.7299 & 0.0041				& 21.8 & 9.7					& 0.19 & 0.06						& 0.085 & 0.027			& (a) \\
\SXV		& r			& 5.0387				& 5.0374 & 0.0015				& 13.8 & 4.0					& 0.63 & 0.13						& 0.177 & 0.037			& (a) \\
		& i			& 5.0665				& \multicolumn{2}{c}{5.0652}			& \multicolumn{2}{c}{13.9}			& 0.14 & 0.08						& 0.038 & 0.022			& (a) \\
		& f			& 5.1015				& \multicolumn{2}{c}{5.1002}			& \multicolumn{2}{c}{14.0}			& 0.45 & 0.11						& 0.113 & 0.029			& (a) \\
\SiXIV		& \Lyalpha		& 6.1804				& 6.1766 & 0.0019				& 12.1 & 5.3					& 0.16 & 0.03						& 0.098 & 0.020			& (a) \\
\SiXIII		& r			& 6.6479				& 6.6421 & 0.0013				& 19.8 & 3.6					& 0.58 & 0.07						& 0.372 & 0.048			& (c) \\
		& i			& 6.6882				& \multicolumn{2}{c}{6.6824}			& \multicolumn{2}{c}{19.9}			& 0.05 & 0.03						& 0.035 & 0.022			& (c) \\
		& f			& 6.7403				& \multicolumn{2}{c}{6.7344}			& \multicolumn{2}{c}{20.1}			& 0.36 & 0.06						& 0.248 & 0.039			& (c) \\
\hline
\multicolumn{10}{c}{CXO$_{030926}$} \\
\hline
\SXVI		& \Lyalpha		& 4.7274				& 4.7291 & 0.0020				&  8.0 & 7.0					& 0.50 & 0.11						& 0.048 & 0.011			& (a) \\
\SXV		& r			& 5.0387				& 5.0370 & 0.0012				& 11.9 & 3.4					& 1.09 & 0.19						& 0.167 & 0.030			& (a) \\
		& i			& 5.0665				& \multicolumn{2}{c}{5.0648}			& \multicolumn{2}{c}{12.0}			& 0.48 & 0.15						& 0.076 & 0.023			& (a) \\
		& f			& 5.1015				& \multicolumn{2}{c}{5.0998}			& \multicolumn{2}{c}{12.0}			& 0.48 & 0.14						& 0.077 & 0.023			& (a) \\
\SiXIV		& \Lyalpha		& 6.1804				& 6.1763 & 0.0032				& 25 & 10					& 0.24 & 0.06						& 0.103 & 0.025			& (a) \\
\SiXIII		& r			& 6.6479				& 6.6452 & 0.0021				& 20.2 & 4.6					& 0.25 & 0.06						& 0.147 & 0.035			& (a) \\
		& i			& 6.6882				& \multicolumn{2}{c}{6.6855}			& \multicolumn{2}{c}{20.3}			& 0.03 & 0.03						& 0.016 & 0.019			& (a) \\
		& f			& 6.7403				& \multicolumn{2}{c}{6.7376}			& \multicolumn{2}{c}{20.5}			& 0.21 & 0.05						& 0.128 & 0.032			& (a) \\
\enddata
\tablecomments{
Values without quoted errors were tied to other fit parameters.
Col.~(3): Rest wavelengths from ATOMDB v1.3.1. For each \Lyalpha\ line we give the
wavelengths of the brighter component; the wavelengths of the fainter components are
4.7328~\angstrom\ (\SXVI) and 6.1858~\angstrom\ (\SiXIV).
Col.~(4): Observed wavelength.
Col.~(5): Observed line width.
Col.~(6): Observed line flux. For each \Lyalpha\ line we give the flux of the brighter component;
the fluxes of the fainter components are half of these values.
Col.~(7): Equivalent width.
Col.~(8): (a)~Fitting to HEG~$\pm 1$ and MEG~$\pm 1$ simultaneously;
(b)~Fitting to HEG~$\pm 1$ and MEG~$\pm 1$ individually and averaging the results.
(c)~Fitting to MEG~$\pm 1$ simultaneously.
}
\end{deluxetable*}

\begin{deluxetable}{lr@{\,$\pm$\,}lr@{\,$\pm$\,}l}
\tabletypesize{\scriptsize}
\tablewidth{0pt}
\tablecaption{Emission Line Shifts And Widths\label{tab:LineVelocities}}
\tablehead{
Ion		& \multicolumn{2}{c}{Shift}	& \multicolumn{2}{c}{Width (FWHM)} \\
		& \multicolumn{2}{c}{(\kmps)}	& \multicolumn{2}{c}{(\kmps)}
}
\startdata
\multicolumn{5}{c}{CXO$_{001119}$} \\
\hline
\SXVI 		&  $-44$ &  76		& \phantom{000}1281 & 311\phantom{000} \\
\SXV		& $-125$ &  42		&  815 & 107 \\
\SiXIV		&  $-34$ &  29		&  800 &  82 \\
\SiXIII		&  $-81$ &  23		&  555 &  59 \\
\hline
\multicolumn{5}{c}{CXO$_{021016}$} \\
\hline
\SXVI 		&  $-76$ &  38		&  894 & 127 \\
\SXV		&  $-77$ &  36		&  809 &  89 \\
\SiXIV		& $-155$ &  24		& 1131 &  78 \\
\SiXIII		&  $-95$ &  18		&  623 &  50 \\
\hline
\multicolumn{5}{c}{CXO$_{030502}$} \\
\hline
\SXVI 		& $-336$ &  38		& 1333 & 121 \\
\SXV		& $-309$ &  54		& 1316 & 119 \\
\SiXIV		& $-514$ &  24		& 1637 &  68 \\
\SiXIII		& $-370$ &  27		& 1111 &  72 \\
\hline
\multicolumn{5}{c}{CXO$_{030616}$} \\
\hline
\SXVI 		& $-387$ &  57		& 1435 & 178 \\
\SXV		& $-446$ &  48		& 1281 & 131 \\
\SiXIV		& $-665$ &  34		& 1376 &  92 \\
\SiXIII		& $-676$ &  41		& 1614 &  99 \\
\hline
\multicolumn{5}{c}{CXO$_{030720}$} \\
\hline
\SXVI 		& $+159$ & 260		& 1382 & 615 \\
\SXV		& $ -77$ &  89		&  821 & 238 \\
\SiXIV		& $-184$ &  92		&  587 & 257 \\
\SiXIII		& $-262$ &  59		&  894 & 162 \\
\hline
\multicolumn{5}{c}{CXO$_{030926}$} \\
\hline
\SXVI 		& $+108$ & 127		&  507 & 444 \\
\SXV		& $-101$ &  71		&  708 & 202 \\
\SiXIV		& $-199$ & 155		& 1213 & 485 \\
\SiXIII		& $-122$ &  95		&  911 & 208
\enddata
\end{deluxetable}

\begin{figure}
\centering
\includegraphics[width=0.9\linewidth,bb=50 50 410 450]{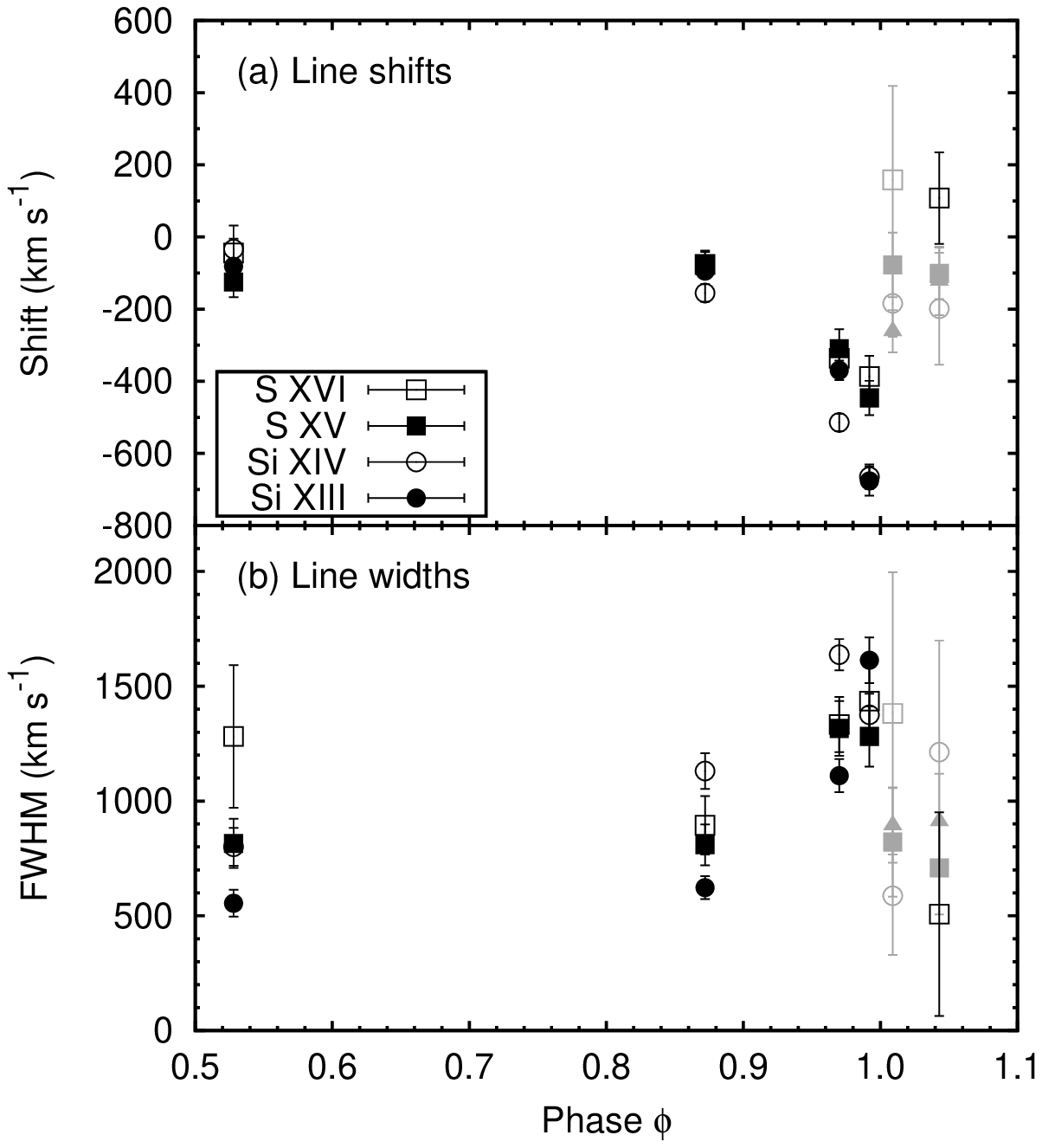}
\caption{Observed emission line shifts (a) and widths (b) plotted against phase. The phase
for each observation is taken from Table~\ref{tab:Observations}. Phase $\phi = 1$ corresponds
to the start of the X-ray minimum in 2003 June. The gray datapoints are for lines
contaminated by the CCE component \citep{hamaguchi07}.\label{fig:ShiftAndWidth}}
\end{figure}

Figure~\ref{fig:ShiftAndWidth} shows
the measured line shifts and widths plotted against phase $\phi$ (see Table~\ref{tab:Observations}),
where $\phi = 1$ corresponds to the start of the X-ray minimum in 2003 June \citep{corcoran05b}.
We have not corrected for the systemic velocity of \ec\ ($-8~\kmps$; \citealp{smith04b}), as
it is negligible compared with the measurement errors.
The general trend of the line shifts is that the lines have small blueshifts of $\sim$100~\kmps\ away from
the X-ray minimum (CXO$_{001119}$ and CXO$_{021016}$), the blueshifts increase to $\sim$300-700~\kmps\ just before
the X-ray minimum (CXO$_{030502}$ and CXO$_{030616}$; note that the lines in CXO$_{030616}$ are generally more blueshifted
than in CXO$_{030502}$), and the blueshifts return to $\sim$100~\kmps\ after the start of X-ray minimum
(CXO$_{030720}$ and CXO$_{030926}$). The exception to this is \SXVI\, which is slightly redshifted in the last two
observations. The general trend of the line widths is that they increase from $\sim$800~\kmps\ (FWHM)
away from the X-ray minimum to $\sim$1400~\kmps\ just before the start of minimum, and then return
to $\sim$800~\kmps\ afterward. However, as noted in \S\ref{sec:Observations}, the last two observations
(being much fainter than the previous ones) are contaminated by the CCE component \citep{hamaguchi07} at
wavelengths longward of about 4\AA. As a result of this contamination, the shifts and widths determined
from the silicon and sulfur lines in these two spectra do not accurately reflect the kinematics of the wind-wind collision
(with the exception of \SXVI\ in CXO$_{030926}$, which is not as badly contaminated).

\begin{figure}
\centering
\includegraphics[width=0.9\linewidth]{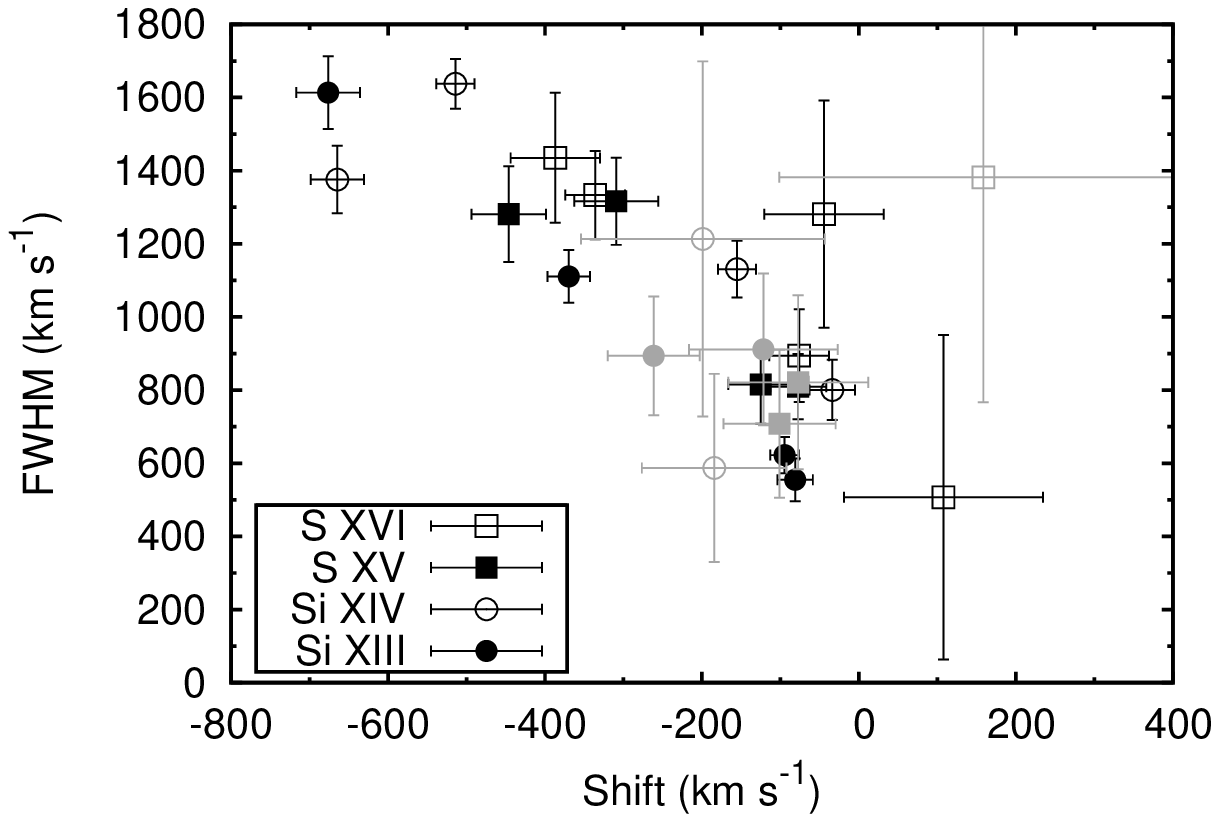}
\caption{The observed emission line widths plotted against the line shifts, showing the correlation between the two.
The gray datapoints are for lines contaminated by the CCE component \citep{hamaguchi07}.
\label{fig:Correlation}}
\end{figure}

The connection between the variation in the line shifts and the variation in the line widths is further
illustrated in Figure~\ref{fig:Correlation}. There is a clear correlation between shift and width, with
the broader lines being more blueshifted. For these data, Spearman's rank correlation coefficient is
$-0.58$, and Kendall's $\tau$ statistic is $-0.47$ \citep{press92}. Both of these statistics show that
correlation is significant at the 1\%\ level.

\begin{figure}
\centering
\includegraphics[width=0.9\linewidth]{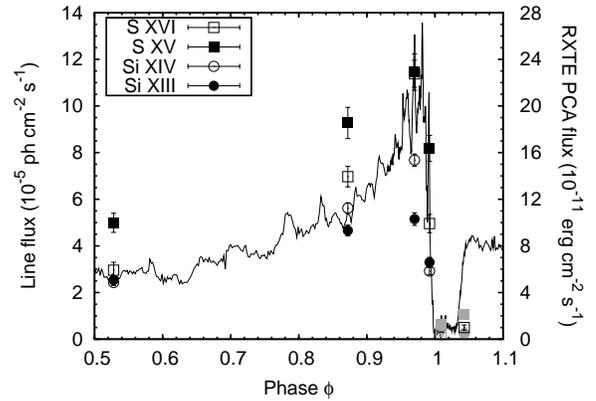}
\caption{Observed emission line fluxes plotted against phase. The phase
for each observation is taken from Table~\ref{tab:Observations}. Phase $\phi = 1$ corresponds
to the start of the X-ray minimum in 2003 June. For \SXV\ and \SiXIII\ we plot the resonance line flux.
The solid line shows the 2--10~\kev\ lightcurve measured with \rxte\ \citep{corcoran05b}.
The gray datapoints are for lines contaminated by the CCE component \citep{hamaguchi07}.
\label{fig:LineFluxes}}
\end{figure}

\begin{figure}
\centering
\includegraphics[width=0.9\linewidth]{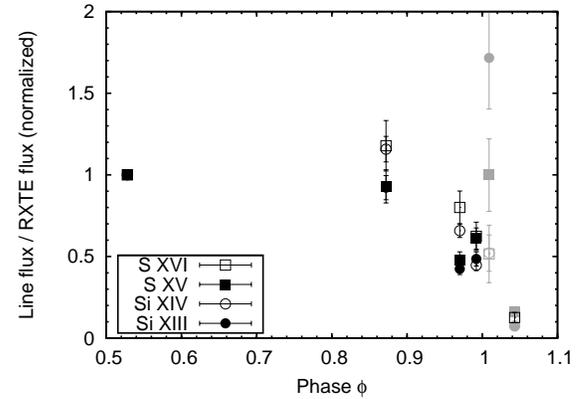}
\caption{Observed emission line fluxes divided by the contemporaneous 2--10~\kev\ flux measured with \rxte\ \citep{corcoran05b}.
For \SXV\ and \SiXIII\ we use the resonance line flux.
For all lines, the flux ratios have been normalized to the values for phase $\phi = 0.528$ (CXO$_{001119}$).
The gray datapoints are for lines contaminated by the CCE component \citep{hamaguchi07}.
\label{fig:LineFluxRatio}}
\end{figure}

Figure~\ref{fig:LineFluxes} shows the variation in the emission line fluxes, plotted with the 2--10~\kev\
\rxte\ lightcurve \citep{corcoran05b}. For \SXV\ and \SiXIII\ we plot the resonance line flux.
As expected, the variation in the line fluxes generally follows
that of the broadband emission. However, not all the lines' fluxes vary in the same way
-- for example, the \SiXIII\ flux does not rise as much as the \SiXIV\ flux in the first three observations,
which in turn does not rise as much as the \SXVI\ flux. These differences between the lines are shown
more clearly in Figure~\ref{fig:LineFluxRatio}, which shows the ratios of the line fluxes to the contemporaneous
2--10~\kev\ flux measured with \rxte\ \citep{corcoran05b}. The ratios are normalized to the values from CXO$_{001119}$.
From CXO$_{001119}$ to CXO$_{021016}$ ($\phi = 0.528$ to 0.872), the emission lines stay fairly constant with respect to the
broadband flux (the \SXVI\ and \SiXIV\ lines actually brighten slightly). However, just before the X-ray minimum (CXO$_{030502}$ and CXO$_{030616}$; $\phi = 0.970$ and 0.992)
the lines grow fainter with respect to the broadband flux. This is what one would expect as the amount of absorption starts increasing:
the emission lines in the $\sim$2--3~\kev\ range will be more strongly attenuated than the broadband flux over the whole 2--10~\kev\ band.
Furthermore, one would expect the longer wavelength lines to show this effect the most. From Figure~\ref{fig:LineFluxRatio} one can
see that in CXO$_{030502}$ this effect is weakest for the \SXVI\ line and strongest for the \SiXIII\ resonance line. Rather surprisingly,
however, the \SiXIV\ line is less affected than the \SXV\ resonance line. Furthermore, the \SXV\ resonance line brightens slightly
with respect to the broadband flux between CXO$_{030502}$ and CXO$_{030616}$.
Note in the final observation, after the recovery (CXO$_{030926}$; $\phi = 1.043$), that the lines are very faint with respect to the broadband flux. This is because
absorption is still having a strong effect on the spectrum, and the observed 2--10~\kev\ flux is coming from shorter wavelengths than the
Si and S lines (one can see from Fig.~\ref{fig:3747spectrum} that most of the flux in CXO$_{030926}$ is shortward of 4~\angstrom).

\subsection{Line Shapes}
\label{subsec:Skew}

\citet{behar07} co-added \Lyalpha, He-like resonance and He-like forbidden lines of Si, S, and Ar
and showed that the resulting profile exhibits a significant asymmetry on the blueward side of the line.
They find that the lines develop blue wings extending to $\sim$2000~\kmps\ in CXO$_{030502}$ and CXO$_{030616}$,
and attribute this to the development of a jet with line-of-sight velocity $\sim$$-2000~\kmps$ near periastron. 
We also looked for evidence of profile asymmetries, using the individual (i.e., non-co-added) lines in each observation. 
A visual inspection of Figures~\ref{fig:SiXIII} and \ref{fig:SiXIV} suggests that some of the lines may indeed be asymmetric.
We find that some of the lines have negative skewness in wavelength (or velocity) space, i.e., an extended tail on the blue side of the line. For example, 
the \SiXIV\ line is skewed in this way in CXO$_{021016}$ and CXO$_{030502}$ (except in the HEG~$+1$ spectrum). However, the observed asymmetry is not as 
apparent in the \SiXIII\ triplet, which makes it difficult to determine whether this apparent asymmetry is real. We
noted that Gaussians give good fits to the individual observed lines. A Gaussian profile would give a bad fit to a strongly skewed line.

In order to quantify the amount of asymmetry in the observed line profiles,
we calculated the skewness of the distribution of photon wavelengths that
make up a given observed line. The skewness $S$ is given by \citep{press92}
\begin{equation}
	S = \frac{1}{N \sigma_\lambda^3} \sum_{i=1}^N (\lambda_\mathrm{i} - \bar{\lambda})^3,
\end{equation}
where $N$ is the number of photons, $\lambda_i$ is the wavelengths of the $i$th photon, and $\bar{\lambda}$ and $\sigma_\lambda$ are
the sample mean and standard deviation of the wavelengths. If our null hypothesis is that the underlying wavelength distribution is Gaussian,
the standard deviation of $S$ is approximately $\sqrt{6/N}$ \citep{press92}.
In the HETGS spectra, the photons are in bins of width 2.5~\mA\ (HEG) and 5~\mA\ (MEG). When estimating $S$, we assume that all the photons in a given
bin have a wavelength equal to the bin's central wavelength. We do not take into account the contribution of the continuum, but
for most observations this should not affect the results too badly, as the lines are much brighter than the continuum.

We looked for skewness in the \Lyalpha\ lines of \SXVI\ and \SiXIV, and the resonance and forbidden lines of \SiXIII. We did not include 
the resonance and forbidden lines of \SXV, as the \SXV\ intercombination line is more prominent (see below), which could affect the results.
In particular we looked for cases where $|S| > 3 \sqrt{6/N}$, although it should be noted that this might not be a strong enough criterion for deciding
if the skewness in the line is significant\footnote{\citet{press92} caution that ``it is good practice to believe in skewnesses only when they are several or many times
as large as [the standard deviation].''}. We examine the individual HEG and MEG~$+1$ and $-1$ orders, and also the co-added first-order HEG and MEG spectra
(to increase the signal-to-noise ratio).  

\begin{figure}
\centering
\includegraphics[width=0.9\linewidth]{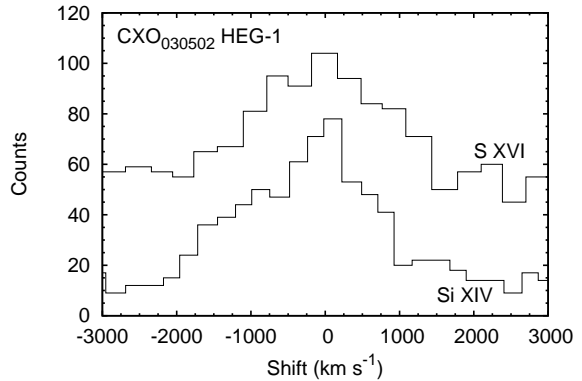}
\caption{Comparison of the HEG~$-1$ profiles of \SXVI\ and \SiXIV\ \Lyalpha\ from CXO$_{030502}$. For clarity, the profiles
have been binned by a factor of two, and the \SXVI\ profile has been shifted upward by 30 counts.
\label{fig:SkewComparison}}
\end{figure}

We found that the \SiXIV\ \Lyalpha\ and \SiXIII\ resonance lines are significantly negatively skewed in CXO$_{030502}$ and CXO$_{030616}$, but the evidence is less
convincing for the forbidden line in these observations (it is significantly skewed in the MEG~$+1$ order, but not in the other orders). The \SXVI\
\Lyalpha\ line is not significantly skewed in these observations (see Fig.~\ref{fig:SkewComparison}, which compares the HEG~$-1$ profiles of \SXVI\ and
\SiXIV\ \Lyalpha\ from CXO$_{030502}$). For the other observations, there is no strong evidence for line skewing -- in a few
cases a line might exhibit skewing in one spectral order, but not in the other three.

Although some of the lines seem to be skewed, visual inspection of Figures~\ref{fig:SiXIII} and \ref{fig:SiXIV} suggests that these asymmetries are relatively modest.
Detailed modeling of these line profile asymmetries reveal finer
details of the wind-wind collision (see \S\ref{sec:SyntheticProfiles}), but the Gaussian-fitting results should provide 
sufficiently accurate information on the gross structure of the wind-wind collision.

We note that, when comparing the results of fitting individual lines from individual orders,
the lines are sometimes slightly offset, possibly due to a slight inaccuracy in the position of the zeroth-order image.
Also, when the lines are analyzed individually, we find that different ions sometimes yield different shifts and widths (see Fig.~\ref{fig:ShiftAndWidth}).
This suggests that adding the profiles from different lines and different spectral orders in order to improve the signal-to-noise \citep{behar07}
might not yield accurate profiles.

\subsection{The $\R$ Ratios of the He-Like Triplets}
\label{subsec:RRatios}

The ratio of the forbidden ($f$) and intercombination ($i$) line intensities of a helium-like ion, $\R = f/i$,
can often provide useful information on the conditions in and location of the emitting plasma.  This is
because the metastable upper level of the forbidden line can be depopulated to the upper level of the intercombination
line by UV photoexcitation or electron collisions: increasing the UV flux or the electron density reduces $\R$
from its low-density, low-UV limit $\R_0$. 
In the case of a hot star possessing a stellar wind, where both electron density and UV flux vary as $1/r^{2}$, $\R<\R_{0}$ implies that the line-emitting region is close to the stellar photosphere. 

In Table~\ref{tab:Rratios} we present the $\R$ ratios for \SiXIII\ and \SXV\ measured from each of our
HETGS spectra. Also in the table we present $\R_0$, calculated by \citet{blumenthal72} at the temperature at
which the triplet has its maximum emissivity (8.9~MK for \SiXIII\ and 14.1~MK for \SXV), and the UV transition
wavelengths to go from the upper level of the forbidden line to the upper levels of the intercombination lines.
In all observations, the \SiXIII\ $\R$ ratio is greater than $\R_0$, implying that the forbidden line is
enhanced with respect to the intercombination line. This has been observed for \OVII\ in the \xmm\ RGS
spectrum of the supernova remnant N132D \citep{behar01}, and for several different ions in the \chandra\ HETGS
spectrum of the WR+O binary WR~140 \citep{pollock05}.

\begin{deluxetable*}{lccccccccc}
\tabletypesize{\scriptsize}
\tablewidth{0pt}
\tablecaption{\R\ Ratios for Helium-like Ions\label{tab:Rratios}}
\tablehead{
		& \colhead{$\lambda_1$\tablenotemark{a}}	& \colhead{$\lambda_2$\tablenotemark{a}}	& \multicolumn{6}{c}{Measured \R\ Ratios} \\
\cline{4-9}
\colhead{Ion}	& \colhead{(\AA)}				& \colhead{(\AA)}				& \colhead{CXO$_{001119}$}	& \colhead{CXO$_{021016}$}	& \colhead{CXO$_{030502}$}	& \colhead{CXO$_{030616}$}	& \colhead{CXO$_{030720}$}	& \colhead{CXO$_{030926}$}	& \colhead{$\R_0$\tablenotemark{b}}	\\
}
\startdata
\SXV		& 738.32					& 673.40					& $2.5 \pm 0.8$			& $1.2 \pm 0.3$			& $1.7 \pm 0.4$			& $1.9 \pm 0.4$			& $3.1 \pm 2.0$			& $1.0 \pm 0.4$			& 2.0					\\
\SiXIII		& 865.14					& 814.69					& $5.0 \pm 1.3$			& $6.2 \pm 1.5$			& $5.7 \pm 1.5$			& $6.8 \pm 2.6$			& $6.9 \pm 4.5$			& $7.6 \pm 8.9$			& 2.5 					\\
\enddata
\tablecomments{$\R = f/i$, where $f$ and $i$ are the forbidden and intercombination line fluxes, respectively.}
\tablenotetext{a}{Wavelengths to go from the upper level of the forbidden line to the upper levels of the intercombination lines -- $\lambda_{1,2}$ are the transition wavelengths
		  for $2\,{^3\mathrm{S}_1} \rightarrow 2\,{^3\mathrm{P}_{1,2}}$, respectively (from CHIANTI; \citealp{dere97,young03}).\\ }
\tablenotetext{b}{Theoretical low-density, low-UV-flux limit at temperature of maximum emissivity (see eq.~[\ref{eq:R}]; values from \citealp{blumenthal72}).\\ }
\end{deluxetable*}

It is possible that too high a continuum level
would lead to line fluxes that are systematically too low. The weak intercombination line would be most severely
affected, and this would lead to an artificially high \R\ ratio. We have investigated whether or not this is
the case in our analysis by adjusting the range of wavelengths we include when fitting to the \SiXIII\ triplet.
The results in Table~\ref{tab:LineResults} were obtained by fitting to the spectra between 6.4 and 7.0~\angstrom\
(note that the plots in Figure~\ref{fig:SiXIII} do not show this full wavelength range). When we use a narrower
range of wavelengths, the forbidden and intercombination fluxes tend to be smaller. While none of the individual
decreases is statistically significant, the fact that there is a systematic shift suggests that with the
narrower wavelength range the line fluxes are systematically underestimated. However, we do not see the opposite
effect when we increase the wavelength range from 6.4--7.0~\angstrom. The amounts by which the fluxes change are
much smaller than when we decreased the wavelength range, and there is no systematic shift in one direction (i.e.,
some fluxes increase slightly, and some decrease slightly). Furthermore, from a visual inspection of the fits,
there is no evidence that a power-law is not a good fit to the continuum over the range of wavelengths that
we use. We have also checked whether we can get a good fit to the spectra with lower values of \R\ by lowering
the normalization of the continuum model by hand from its best-fit value. This should, in principle, increase
the flux of the weak intercombination line relative to that of the stronger forbidden line. However, we find
that lower continuum levels still lead to values of \R\ greater than $\R_0$, and if the continuum normalization
is too low, the fit to the continuum is very poor. From these observations, we conclude that our large \R\ ratios
are probably not due to an inaccurate continuum level.

Because of the rather large
errors on \R, only the \R\ ratios for CXO$_{021016}$ and CXO$_{030502}$ differ by more than $2\sigma$ from $\R_0$,
and no observed \R\ ratio differs by more than $3\sigma$ from $\R_0$. However, if we take as a null hypothesis
that $\R = \R_0$ for all six of our observations, this gives $\chi^2 = 18.35$ for 6 degrees of freedom
($\chi^2$ probability = 0.54\%). This implies that $\R$ is significantly different from $\R_0$ (as calculated
by \citealp{blumenthal72}) for at least some of our spectra. However, the combination of uncertainties in the ratios
and the atomic models prevent us from drawing any strong conclusions.
One might suppose that the large observed values of $\R_0$ are due to inner-shell ionization of Li-like Si to He-like Si
($1\mathrm{s}^2 2\mathrm{s} + e^- \rightarrow 1\mathrm{s} 2\mathrm{s} + 2e^-$), producing ions in the upper level
of the forbidden line. This could arise in an ionizing plasma, because inner-shell ionization requires both a high electron
temperature and an abundance of Li-like ions, two conditions which tend not to hold simultaneously in an equilibrium plasma \citep{mewe78b}.
However, Li-like satellite lines would also become important in an ionizing plasma, so that the empirically observed ratio would no longer
reflect just the ratio of the He-like ions. Without making detailed NEI calculations, it is not safe to make even qualitative predictions for the
expected \R\ ratios of an ionizing plasma.

For \SXV\ the values of $\R$ are generally close to $\R_0$. In an equilibrium plasma, $\R$ can be used to place constraints on the
electron density \Ne\ and the UV flux, and hence place constraints on the location of the X-ray--emitting plasma.
One can express $\R$ as a function of \Ne\ and the photoexcitation rate $\phi$ to go from the upper level of the
forbidden line to the upper level of the intercombination line:
\begin{equation}
	\R = \frac{\R_0}{1 + (\phi / \phic) + (\Ne / \nc)},
\label{eq:R}
\end{equation}
where \phic\ and \nc\ are quantities dependent only on atomic parameters and the electron temperature \citep{blumenthal72}.
The ratio \R\ tends toward the limit $\R_0$ when $\phi \ll \phic$ and $\Ne \ll \nc$. \citet{blumenthal72} give
$\nc = 1.9 \times 10^{14}$~\pcc\ and $\phic = 9.16 \times 10^5$~\ps\ for \SXV\ at the temperature of maximum emissivity.
If we assume $\Mdotc  = 10^{-5}$~\Msolpy\ and $\vc = 3000$~\kmps\ for the companion \citep{pittard02a}, we find
that $\Ne \ll \nc$ everywhere in the companion's wind; unless the shock compression ratio is very large
(several hundred or more), this will also be true in the wind-wind collision region. Thus, electron
collisions are not expected to affect the $\R$ ratio.

\citet{blumenthal72} also tabulate $\phistar / \phic$,
where $\phistar$ is the photoexcitation rate on the surface of a $10^5$-K blackbody. We estimate $\phistar / \phic$
for \ec's companion by scaling the \citet{blumenthal72} value for \SXV, assuming the companion is
a blackbody with $T = \mbox{36,000}$~K (this is in the middle of the range of effective temperatures given by \citealp{verner05}: $\mbox{34,000}~\K \le T_\mathrm{eff} \le \mbox{38,000}~\K$).
We obtain $\phistar / \phic = 1.6$, which means that on the surface of the companion we have $\R = \R_0 / (1 + 1.6) = 0.77$,
with $\R$ increasing toward $\R_0$ as we move away from the star. 
Unfortunately, the fact that $\R$ is fairly large even on the surface of the companion, and the large errors on $\R$ in Table~\ref{tab:Rratios},
make it difficult to place strong constraints on the location of the emitting plasma. As one moves away from the
companion, the photoexcitation rate decreases as $\phi(r)/\phic = 2 W(r) \phistar/\phic$, where $W(r) = 0.5[1-\sqrt{1-(\rc/r)^2}]$ is the
geometrical dilution factor, and $r$ is the distance from the center of the companion, whose radius is \rc.
Note that \rc\ is not well known, though \citet{ishibashi99} estimate $\rc \sim 50 \Rsol$. If we take the
result for CXO$_{001119}$, and say that the measurements imply $\R > 1.7$ (i.e., the $1\sigma$ lower limit), this gives
$r > 1.6 \rc$ for the location of the X-ray--emitting plasma.

\begin{figure}
\centering
\includegraphics[width=0.9\linewidth]{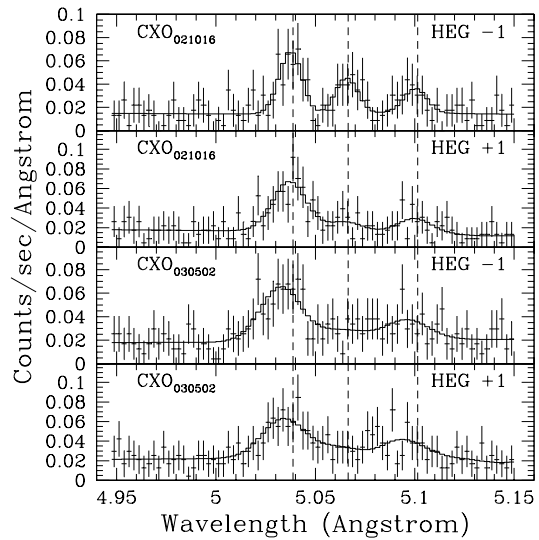}
\caption{First-order HEG spectra of the \SXV\ f-i-r triplet from CXO$_{021016}$ (\textit{top two panels}) and CXO$_{030502}$ (\textit{bottom two panels}).
The histograms show the best-fitting Gaussian line models obtained by fitting to each spectral order individually.
The vertical dashed lines show the rest wavelengths of the resonance, intercombination, and forbidden lines (5.0387, 5.0665, and 5.1015~\angstrom, respectively).
\label{fig:SXV}}
\end{figure}

The low \SXV\ \R\ ratio for CXO$_{021016}$ seems to suggest that the \SXV\ emission originates close to the companion in that observation.
If we were to take $\R < 1.5$, this would imply $r < 1.2 \rc$. However, closer inspection of the spectra shows that the \SXV\ intercombination
line is noticeably brighter in the HEG~$-1$ spectrum than in the HEG~$+1$ spectrum.
This can be seen in Figure~\ref{fig:SXV}, which shows the \SXV\ triplet from the CXO$_{021016}$
HEG spectra, along with the \SXV\ triplet from the CXO$_{030502}$ HEG spectra for comparison. The \R\ ratios for CXO$_{021016}$ from
the individual HEG orders are $0.6 \pm 0.2$ ($-1$) and $1.9 \pm 1.1$ ($+1$), while the \R\ ratio obtained from the HEG~$+1$ fit results combined
with those from the two MEG orders is $1.8 \pm 0.5$. It therefore seems that the low \R\ ratio for CXO$_{021016}$ is mainly due to the bright
intercombination line in the HEG~$-1$ spectrum. We have examined the first-order HEG image, using the CIAO tool \texttt{tg\_scale\_reg} to
establish the position of the \SXV\ intercombination line. We find that there is no detector feature (such as a hot pixel) or X-ray source
which is contaminating the intercombination line in the HEG~$-1$ spectrum. We have also compared the forbidden and intercombination line
fluxes measured in the $+1$ and $-1$ orders of both gratings for each observation. We have done this for \SiXIII\ and \SXV. In principle,
this would be a total of 48 comparisons (6~observations $\times$ 2~gratings $\times$ 2~ions $\times$ 2~lines). However, as we cannot fit to
individual orders in all cases, in practice we find we can only make 30 such comparisons. Among these comparisons, only the CXO$_{021016}$
\SXV\ intercombination line measured by the HEG differs by more than $2\sigma$ between the $+1$ and $-1$ orders (the difference is $2.2\sigma$).
With the null hypothesis that the line flux is the same for both orders, the probability of such a large difference is 2.8\%. Therefore,
it is not surprising that, among our set of 30 comparisons between the positive and negative first-order spectra, we find one case where
the two values differ by $2.2\sigma$. This suggests that the large intercombination line flux in the CXO$_{021016}$ HEG~$-1$ spectrum is
a statistical fluke, and that there is no convincing evidence that the \SXV\ \R\ ratio for this observation really is significantly lower
than those in the other observations.

As a final point, it should be noted that the spectral type of the companion is not known, and a 36,000-K blackbody may poorly represent its
UV flux at the wavelengths relevant to the above analysis. A more detailed model of its spectrum is required to place more accurate constraints
on the location of the X-ray--emitting plasma.

\subsection{The $G$ Ratios of the He-Like Triplets, and the Ratios of H-like to He-like Lines}
\label{subsec:GRatios}

\begin{figure}
\centering
\includegraphics[width=0.48\linewidth, bb=50 110 410 914]{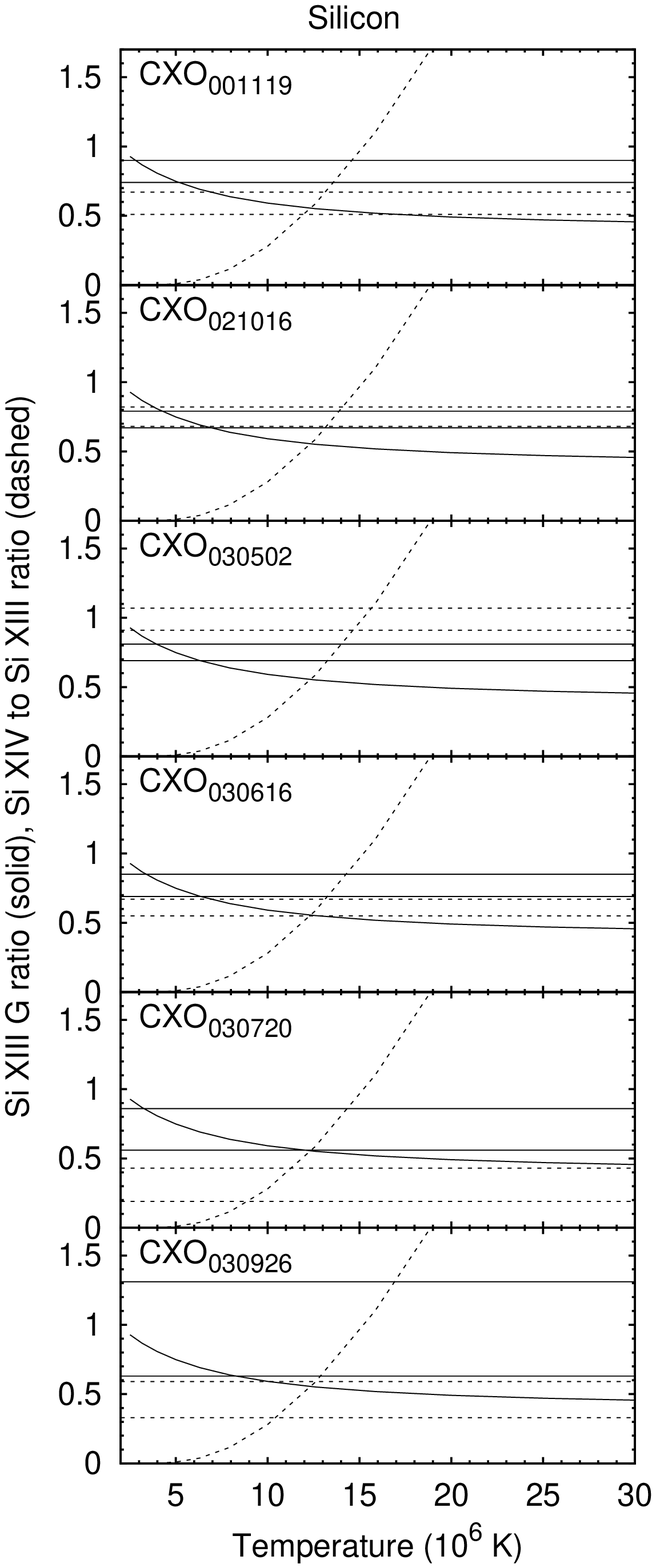}
\includegraphics[width=0.48\linewidth, bb=50 110 410 914]{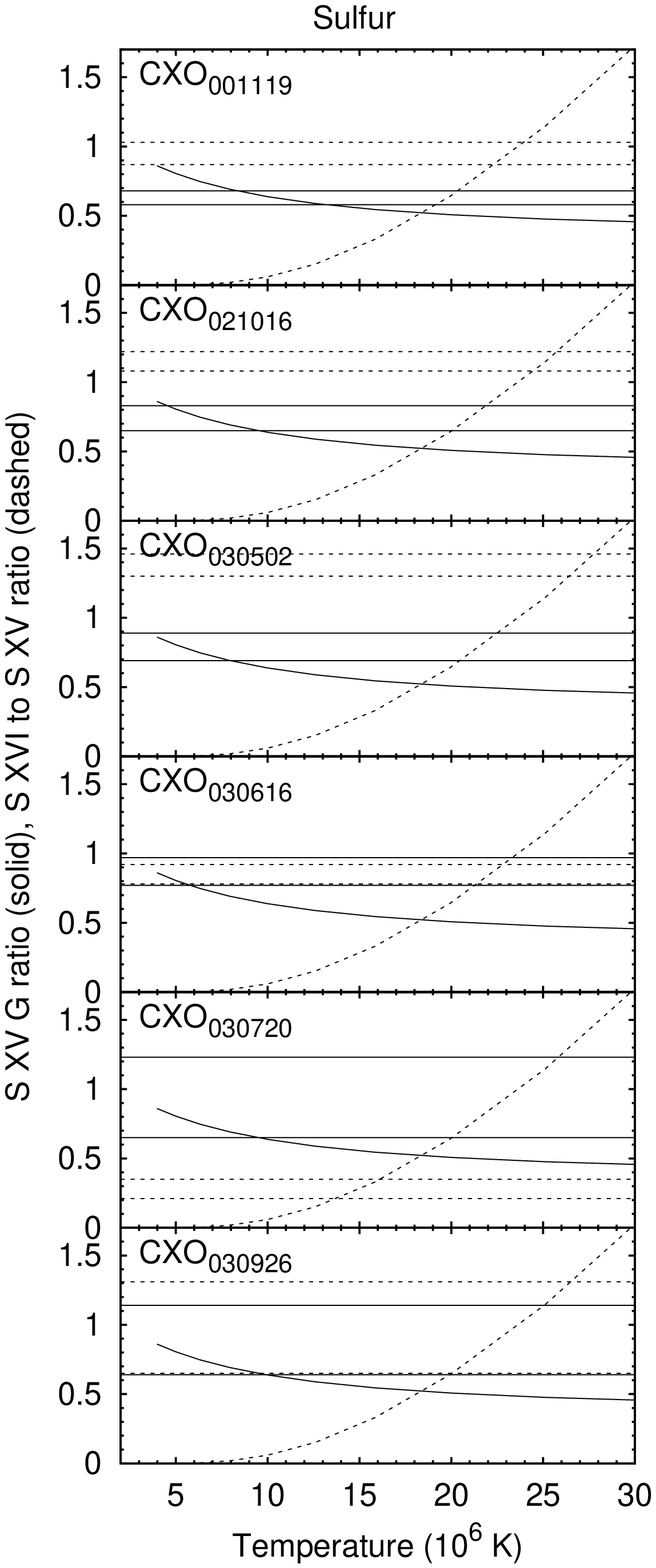}
\caption{$G=\{(\mathrm{forbidden}+\mathrm{intercombination})/\mathrm{resonance}\}$ line flux ratios and H-like to He-like line flux ratios for
silicon (\textit{left}) and sulfur (\textit{right}). The horizontal solid lines show the observed values of $G$ ($1\sigma$ limits),
while the curved solid lines show the theoretical values as a function of temperature from APED. The dashed lines show similar information
for the flux ratio of the H-like \Lyalpha\ line to the He-like resonance line. (For the latter ratio we use the flux of the brighter component
of the \Lyalpha\ doublet.)\label{fig:Gratio}}
\end{figure}

We also measured the $G=(f+i)/r$ line ratios for the helium-like \SiXIII\ and \SXV\ triplets. The $G$ ratio decreases with temperature.
It is also sensitive to densities for $n>10^{12}~\pcc$, which is well above the range of densities expected in the wind-wind collision
in $\eta$~Car from hydrodynamical simulations ($<10^{11}~\pcc$; see for example \citealp{pittard02a}).  The measured $G$ ratios are given
in Table~\ref{tab:Gratios}, and are plotted in Figure~\ref{fig:Gratio} (horizontal solid lines). Also plotted in this figure is the temperature
dependence of the $G$ ratios based from the Astrophysical Plasma Emission Database (APED; \citealp{smith01a}), version 1.3.1 (curved solid lines).
The measured $G$ ratios generally imply temperatures of $T < 8 \times 10^6~\K$ for \SiXIII\ and $T < 13 \times 10^6~\K$ for \SXV.

\begin{deluxetable*}{lcccccc}
\tabletypesize{\scriptsize}
\tablewidth{0pt}
\tablecaption{$G$ Ratios for Helium-like Ions\label{tab:Gratios}}
\tablehead{
\colhead{Ion}	& \colhead{CXO$_{001119}$}	& \colhead{CXO$_{021016}$}	& \colhead{CXO$_{030502}$}	& \colhead{CXO$_{030616}$}	& \colhead{CXO$_{030720}$}	& \colhead{CXO$_{030926}$}	\\
}
\startdata
\SXV		& $0.68 \pm 0.10$		& $0.74 \pm 0.09$		& $0.79 \pm 0.10$		& $0.87 \pm 0.10$		& $0.94 \pm 0.29$		& $0.89 \pm 0.25$		\\
\SiXIII		& $0.82 \pm 0.08$		& $0.73 \pm 0.06$		& $0.75 \pm 0.06$		& $0.77 \pm 0.08$		& $0.71 \pm 0.15$		& $0.97 \pm 0.34$		\\
\enddata
\tablecomments{$G = (f+i)/r$, where $f$, $i$, and $r$ are the forbidden, intercombination, and resonance line fluxes, respectively.}
\end{deluxetable*}

Also shown in Figure~\ref{fig:Gratio} is the ratio of the flux of the brighter component of the H-like \Lyalpha\ line to that of the He-like resonance
line -- the horizontal dashed lines show the observed values, while the curved dashed lines show the theoretical values, also from APED.
This ratio is an increasing function of temperature, as the ionization balance shifts from He-like ions to H-like ions. The observed
ratios generally imply temperatures of $T \sim 12$--$15 \times 10^6~\K$ for silicon and $T \sim 20$--$25 \times 10^6~\K$ for sulfur.

With the exception of the silicon measurements from CXO$_{030720}$ (which was obtained during the X-ray minimum, and has much lower signal-to-noise
than the other observations), the temperatures implied by the $G$ ratios are lower than those implied by the H-like to He-like line flux ratios.
This could be taken to imply that the gas is out of equilibrium, as the electron temperature (given by the $G$ ratio) is lower
than the ionization temperature (given by the H-like to He-like flux ratio), suggesting that the plasma is overionized. However, an
alternative explanation is that the H-like and He-like line emission originate from different regions of the wind-wind collision, with
the He-like emission originating from a region with a lower temperature than the H-like emission. This is what one would expect from a plasma
with a range of temperatures (such as a wind-wind collision, where the shocked gas near the stagnation point is hotter than gas further out).
It is also possible that absorption in the cool, unshocked winds of the stars is affecting the H-like to He-like line flux ratios. As the
He-like line from a given element is at a lower energy than the H-like line, it will be more strongly absorbed by the cool, unshocked stellar
winds. This would tend to increase the temperature inferred from the observed H-like to He-like line flux ratio.

The main conclusion of this section and the previous section is that we have not found unambiguous evidence of non-equilibrium conditions
from the observed line flux ratios. However, when we compare the observed line profiles with those predicted by a model of the wind-wind
collision, we find that the emitting region is much smaller than expected if the wind-wind collision were in equilibrium, suggesting that
the wind-wind collision may be out of equilibrium. This modeling is described in \S\ref{sec:SyntheticProfiles}, and the results are
discussed in \S\ref{subsec:ConstrainingInteractionRegion}.

\section{A SIMPLE GEOMETRICAL MODEL OF THE COLLIDING WIND REGION}
\label{sec:GeometricalModel}

It is clear from the preceding section that \ec's X-ray emission lines show variability around the time of
the X-ray minimum. We first attempt to understand this variability in terms of a simple geometrical characterization of the emission region as a conical surface of constant opening angle.
This analysis has been applied to features in optical emission lines from WR~79 to constrain orbital and other parameters of the system \citep{luhrs97},
and also to X-ray emission lines from WR~140 \citep{pollock05} and $\gamma^{2}$ Velorum \citep{henley05a}.

\subsection{Description of the Model}
\label{subsec:ModelDescription}

\begin{figure}
\centering
\includegraphics[width=0.9\linewidth]{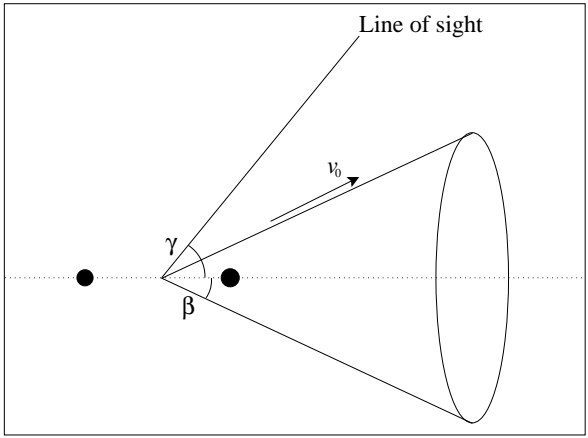}
\caption{Geometrical model of the wind-wind collision in \ec. The solid circles represent the two
stars. The cone (with opening half-angle $\beta$) represents the wind-wind interaction region (along which X-ray--emitting
material is streaming at speed $v_0$). The viewing angle $\gamma$ is the angle between the line of sight and the
line of centers.}
\label{fig:GeometricalModel}
\end{figure}

We assume that the X-ray emission comes from a conical emission region with opening half-angle $\beta$,
whose symmetry axis lies along the line of centers with the apex pointing toward the primary star, and along
which material streams at speed $v_0$. The viewing angle $\gamma$ is the
angle between the line of centers and the line of sight. The geometry is illustrated in
Figure~\ref{fig:GeometricalModel}. Assuming that there is no azimuthal velocity component, the centroid shift
($\bar{v}$) and velocity range ($v_\mathrm{max} - v_\mathrm{min}$) of an emission line are given by \citep{luhrs97,pollock05,henley05a}
\begin{eqnarray}
        \bar{v}       &=& -v_0 \cos \beta \cos \gamma,
\label{eq:Shift} \\
        v_\mathrm{max} - v_\mathrm{min} &=& 2 \Delta v = 2 v_0 \sin \beta \sin \gamma.
\label{eq:Width}
\end{eqnarray}
The viewing angle $\gamma$ can be calculated from the orbital solution. We first define $\Psi$ as the angle between the line of centers
at the time being considered and the line of centers when the companion star is in front; $\Psi$ can be calculated from the
true anomaly $\Phi$ and the longitude of periastron $\omega$:
\begin{equation}
	\cos \Psi = \cos ( \Phi + \omega - 90\degr ).
\end{equation}
If $i$ is the orbital inclination, then $\gamma$ is given by
\begin{equation}
	\cos \gamma = \cos \Psi \sin i.
\end{equation}

When comparing the predictions of this model with the observed data, $\bar{v}$ may simply be equated to the shifts in
Table~\ref{tab:LineVelocities}. The relation between $\Delta v$ and the measured Gaussian line widths is less straightforward.
We assume that the observed velocities
range from $v_\mathrm{min} \approx (\mathrm{Shift}) - (\mathrm{FWHM})$ to $v_\mathrm{max} \approx (\mathrm{Shift}) + (\mathrm{FWHM})$,
and proceed by simply equating $\Delta v$ in equation~(\ref{eq:Width}) to the observed FWHM.

\begin{figure}
\centering
\includegraphics[width=0.9\linewidth]{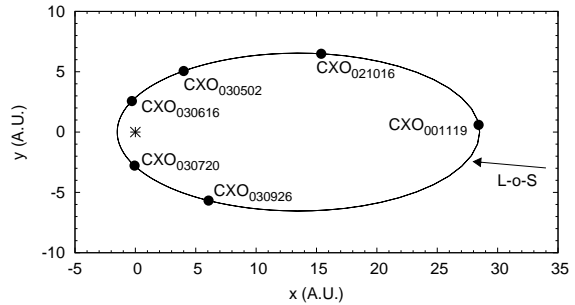}
\caption{The orbit of \ec's companion plotted in the rest frame of the primary (marked with an asterisk at the origin).
The positions of the companion at the times of the \chandra\ HETGS observations are marked with black circles.
The arrow shows the line of sight (L-of-S) projected into the orbital plane.
This orbit was plotted using the parameters in Table~\ref{tab:Orbit}. The length scale of the orbit is set by assuming masses of
80\Msol\ and 30\Msol\ for the primary and the companion, respectively \citep{corcoran01a}.\label{fig:Orbit}}
\end{figure}

The orbital parameters we assume initially are given in Table~\ref{tab:Orbit},
which are largely based upon \citepossessive{corcoran01a} analysis of the \rxte\ light curve, with a revised
period from \citet{corcoran05b}. Note that the time of periastron passage $T_0$ in Table~\ref{tab:Orbit} is actually
the time of the start of the X-ray minimum \citep{corcoran05b}, which was used to calculate the phases in
Table~\ref{tab:Observations}. However, as periastron is expected to occur
near the time of the X-ray minimum,
assuming the two times are equal has little effect on the results. If the time of periastron passage is allowed to differ
from the time of the start of the X-ray minimum, this will only result in the curves calculated below being shifted slightly
to the left or right. The orbit specified by the parameters in Table~\ref{tab:Orbit} is shown in Figure~\ref{fig:Orbit}.
The length scale of the orbit is set by assuming masses of 80\Msol\ and 30\Msol\ for the primary and the companion, respectively
\citep{corcoran01a}. However, the scale of the orbit is not important for our analysis -- all that matters is how the viewing
angle varies with time.

\begin{deluxetable}{lcc}
\tablewidth{0pt}
\tablecaption{$\eta$ Carinae Orbital Parameters\label{tab:Orbit}}
\tablehead{
\colhead{Parameter}			& \colhead{Value}	& \colhead{Reference}
}
\startdata
$T_0$ (periastron) (JD)			& 2\,450\,799.792	& (1) \\
Period $P$ (d)				& 2024			& (1) \\
Eccentricity $e$			& 0.90			& (2) \\
Longitude of periastron $\omega$ (deg)	& 275			& (2) \\
Inclination $i$ (deg)			& 50			& (2) \\
\enddata
\tablecomments{
(1) \citet{corcoran05b}; (2) \citet{corcoran01a}.
}
\end{deluxetable}

In addition to the orbital elements, we also need to assume an opening angle $\beta$ for the wind-wind interaction region,
and a speed $v_0$ for material streaming along the cone. From hydrodynamical simulations of the wind-wind collision in \ec,
we adopt a shock opening half-angle $\beta = 58 \degr$ \citep{henley05b}. This is consistent with the shock opening angle
estimated from the equivalent width of the Fe fluorescence line measured with \xmm\ \citep{hamaguchi07}. At large distances from the
line of centers, the velocity along the shock cone $v_0$ tends toward the terminal velocity of the companion star's
wind (3000~\kmps). However, the observed emission lines are likely to originate from nearer to the line of centers
\citep{henley03} -- the outer regions are not favored for X-ray line emission because (a) the gas number density $n$ falls
off with distance from the line of centers (and the line luminosity scales as $n^2$) and (b) the gas temperature also falls off,
reducing the populations of H- and He-like ions whose lines we are discussing here. However, very near the line of
centers (where $v_0$ is much lower), the gas is too hot for most of the observed ions to exist in significant amounts,
and so the line emission falls off here too despite the greatly increased density. 
Using the line profile model described
in \citet{henley03}, we find that most of the line emission should originate where $v_0 \approx 2000$--3000~\kmps.

\begin{figure}
\centering
\includegraphics[width=0.9\linewidth,bb=50 50 410 450]{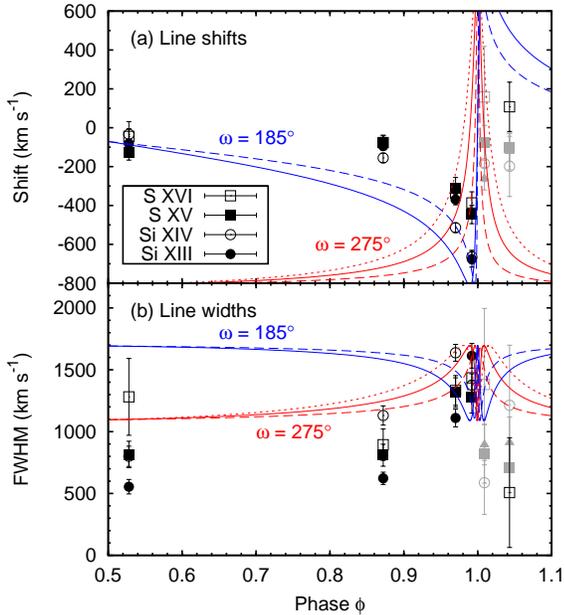}
\caption{Comparison of the predictions of the simple geometrical model (\S\ref{sec:GeometricalModel})
with the observed line shifts and widths. The solid red curves were generated from the orbital parameters in
Table~\ref{tab:Orbit} ($i = 50\degr$, $\omega = 275\degr$, $e = 0.9$). The other red curves show the
effect of varying the eccentricity: $e = 0.85$ (\textit{short dashed}) and $e = 0.95$ (\textit{long dashed}).
The blue curves show the effect of adjusting $\omega$ to 185\degr, for two different eccentricities:
$e = 0.9$ (\textit{solid}) and $e = 0.95$ (\textit{long dashed}).
The gray datapoints are for lines contaminated by the CCE component \citep{hamaguchi07}.
\label{fig:ShiftAndWidthGeometricalModel}}
\end{figure}

The solid red line in Figure~\ref{fig:ShiftAndWidthGeometricalModel} shows the results of the geometrical model
compared to the observed line shifts and widths. The observed variation in the
line widths is in qualitative agreement with the model
in that the widths increase around $\phi = 1$ and
decrease again afterward, although the model parameters we have used predict larger widths than are observed. However,
the agreement between the observed and model velocity shifts is poor using the model parameters adopted above:
away from the X-ray minimum, 
the model predicts large
blueshifts of $\sim$800~\kmps, whereas we observe much smaller blueshifts of $\sim$100~\kmps, while near
X-ray minimum, the model predicts redshifted lines, in contrast to the increasing blueshifts which we observe. 
Some of this discrepancy may be due to the assumed values of the shock parameters and orbital elements.  We consider the dependence of the model velocities and widths on 
the parameters $v_0$, $\beta$, $i$, $\omega$ and $e$ below.

\subsection{Dependence on the Shock Parameters}

The model
line centroids and widths depend on the conditions assumed for the boundary surface of the idealized wind-wind interaction, namely the flow speed $v_{0}$ and the cone opening angle $\beta$.  Since the flow speed $v_0$ appears as a multiplicative constant in equations~(\ref{eq:Shift}) and (\ref{eq:Width}), varying $v_0$ simply
varies the amplitude of the variation in the predicted shifts and widths. For example, lowering $v_0$ by a few hundred \kmps\
would bring the predicted widths into better agreement with the observed widths. However, the discrepancy between the predicted
and observed shifts 
would 
still exist.

From inspection of equations~(\ref{eq:Shift}) and (\ref{eq:Width}), one can see that varying $\beta$ will also
change the amplitude
of the variation in the predicted shifts and widths. In particular, increasing $\beta$ decreases the amplitude of the 
velocity shift and increases the amplitude of the variation in line width.
However, if everything else is kept the same, the model still predicts redshifted
lines around the time of the X-ray minimum, instead of the observed blueshifted lines.

\subsection{Dependence on the Orbital Elements}

Varying the inclination $i$ varies the amplitude of the variation in the viewing angle $\gamma$.
In an edge-on binary ($i = 90\degr$), $\gamma$ varies from 0\degr\ at one
conjunction\footnote{Depending on which star has the more powerful wind.}, to 90\degr\ at quadrature, to 180\degr\ at the other conjunction,
and back again to 0\degr. In contrast, a face-on binary ($i = 0\degr$) is always observed at $\gamma = 90 \degr$. In general, $\gamma$
varies between $90\degr - i$ and $90\degr + i$ during the course of the orbit.
The result of this is that varying the inclination $i$ also varies the amplitude of the shift and width variations.
Maximum variability occurs when $i = 90\degr$, and there is no variability for $i = 0\degr$.
However, whereas reducing $v_0$ or $\beta$ reduces the predicted widths as well as the amplitude of the variation, as $i$ tends to 0\degr\ 
the width tends to $v_0 \sin \beta$ rather than to zero (see eq.~[\ref{eq:Width}]).
We find that simply varying the inclination cannot bring the model into good agreement with the observations.

We also considered the effect of changing the orbital eccentricity.  Increasing the eccentricity means that the viewing angle changes
more rapidly during periastron passage. This in turn means that
the predicted shifts and widths will change more rapidly. As a result, the peak at $\phi = 1$ in the solid red curve in
Figure~\ref{fig:ShiftAndWidthGeometricalModel}(a) and the double-peaked feature at $\phi = 1$ in the solid red curve in
Figure~\ref{fig:ShiftAndWidthGeometricalModel}(b) both become narrower with increasing eccentricity, and broader
with decreasing eccentricity. This is shown by the dashed red curves in Figure~\ref{fig:ShiftAndWidthGeometricalModel}.

Finally, varying the longitude of periastron $\omega$
has the largest effect on determining the phase dependence of the velocities in the model.
The solid blue curves in Figure~\ref{fig:ShiftAndWidthGeometricalModel} show a model
with $\omega = 185\degr$, which means that the orbit
has been rotated 90\degr\ clockwise. In this orientation the semimajor
axis is approximately perpendicular to the line of sight, and the companion passes in front of the primary just before periastron.
One can see that this $\omega$ does yield lines with small shifts away
from $\phi = 1$, and with increasing blueshifts as $\phi$ approaches 1. However, the increase in the model blueshift
occurs too soon in phase compared with the observed centroid shifts. Increasing the eccentricity helps by delaying the blueshift in phase,
and by making the change in centroid velocity more rapid near periastron passage. The blue dashed curve in
Figure~\ref{fig:ShiftAndWidthGeometricalModel}(a) shows a model in which $\omega = 185\degr$ and $e = 0.95$ instead of 0.9. Although the
agreement is not formally acceptable, this model is in rough qualitative agreement with the variation in the line shifts prior to the
X-ray minimum, though it fails to describe the observed variations in line widths.  Further adjustment of $e, \omega$, and $v_{o}$ or $\beta$ might further improve the agreement.

After the X-ray minimum, the new model predicts lines redshifted by a few hundred \kmps, whereas the observed lines generally
have small ($\sim$100~\kmps) blueshifts. As noted in \S\ref{sec:Observations},
the silicon and sulfur lines are significantly contaminated by emission from the CCE component in the last two \chandra\ spectra (except for \SXVI\ in CXO$_{030926}$).
This means that these lines do not accurately reflect the centroids of the lines produced by the wind-wind collision.
However, with this new value of $\omega$
the agreement between the predicted and observed widths shown in Figure~\ref{fig:ShiftAndWidthGeometricalModel}(b) is poorer than it was for the original model: away from the X-ray
minimum the new model predicts larger widths than are observed, and the predicted widths decrease around $\phi = 1$, instead
of increasing.

In summary, we have shown how adjusting the various parameters in our geometrical model for the line shifts and widths affects the model
predictions. We find that by adjusting certain parameters it is possible to bring the model into rough
qualitative agreement with the observations for a subset of the shifts or widths, but we have not found a set of parameters which describes 
both the line shifts and variations in line widths simultaneously in all of the observations well, though admittedly 
we have not carried out a complete exploration of the whole parameter space. However, by seeing how the
individual model parameters affect the model curves it is not easy to see which combination of parameters 
would bring this simple geometrical model into good agreement with the observations.

\section{SYNTHETIC LINE PROFILE MODELING}
\label{sec:SyntheticProfiles}

In the previous section we showed that there is poor agreement between the shifts and widths predicted by the simple geometrical
model, and those that are observed in the HETGS spectra of \ec. With a longitude of periastron $\omega \approx 270 \degr$,
we can get reasonable agreement with the observed variation of the widths, and with $\omega \approx 180 \degr$
we can get reasonable agreement with the observed variation of the shifts. However, we cannot match the variation of both
simultaneously. Furthermore, when the axis of shock cone is nearly perpendicular to the line of sight (i.e., $\gamma \approx 90\degr$),
the above-described model predicts broad double-peaked line profiles (with the peaks at $\sim \pm v_0 \sin \beta$). With
the orbital parameters discussed above, we expect at least one of our observations to have $\gamma \approx 90\degr$.
However, we do not see double-peaked profiles in any of our spectra. To address these issues, we have developed a more sophisticated
model for calculating emission line profiles, taking into account both the shape of the wind-wind collision region and the variation
in the speed at which material flows away from the stagnation point.  \citet{falceta06} showed that a similar detailed line profile model, including turbulent broadening and intrinsic absorption was needed to fit the phase-dependent, asymmetric \ion{C}{3} 5696~\AA\ line from the WR+O colliding wind binary Br22.

\subsection{Description of the Model}

\begin{figure}
\centering
\includegraphics[width=0.9\linewidth, bb=80 50 380 302]{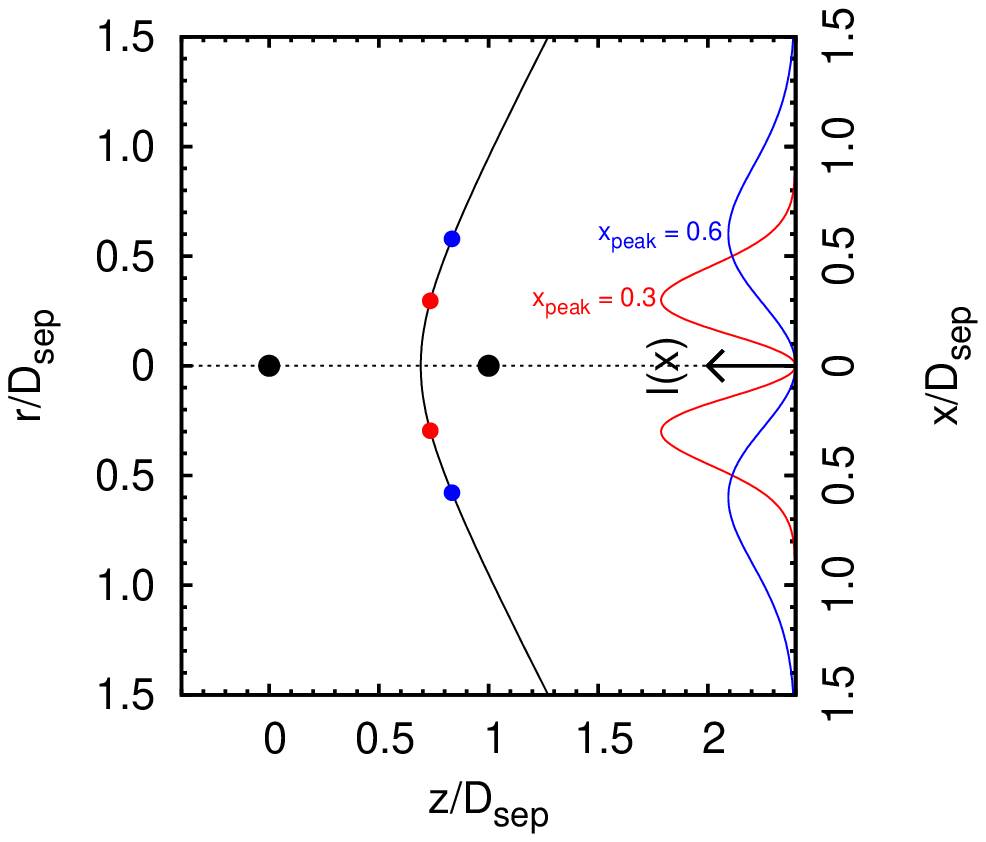}
\caption{The shape of the wind-wind collision surface in \ec, calculated using the equations of \citet{canto96}. The two stars
are shown by the black circles on the $z$ axis: the primary is at the origin, and the companion is at (1,0).
Distances are normalized to the stellar separation \Dsep. The curves to the right show $l(x)$ (see eq.~[\ref{eq:l(x)}]) plotted for two different values of \xpeak\
(\textit{red}: $\xpeak = 0.3$; \textit{blue}: $\xpeak = 0.6$). Note that $x$ is actually the distance measured
along the wind-wind collision surface. The points at $x=0.3$ and $x=0.6$ are marked on the surface with the red and blue circles.
\label{fig:CantoSurface}}
\end{figure}

We calculate the shape of the wind-wind collision region using the results of \citet{canto96}, who have derived equations for the
surface of momentum balance between two colliding spherical winds. This model is for two totally radiative winds with complete
mixing between them. While this is not expected to be the case in \ec, it provides a useful starting point for modeling
the line emission, in particular for determining the shape of the surface of momentum balance.
We assume that the X-ray--emitting region is optically and spatially thin, and coincident with the surface of momentum balance.
The shape of the wind-wind collision surface depends on the wind momentum ratio $\eta = \Mdotc \vc / \Mdoteta \veta$, and the flow
speed along the surface also depends on the wind speeds of the stars \vc\ and \veta. For our canonical model we adopt
$\eta = 0.2$, $\vc = 3000~\kmps$ and $\veta = 500~\kmps$ \citep{pittard02a}. The resulting shape of the wind-wind collision
surface is shown in Figure~\ref{fig:CantoSurface}.

Using the \citet{canto96} equations, we find that the flow speed along the wind-wind collision surface away from the stagnation
point tends toward $\sim$900~\kmps\ far from the stagnation point. However, hydrodynamical simulations suggest that
the flow speed in the X-ray--emitting region tends toward the wind speed of the companion (i.e., $\sim$3000~\kmps). To allow for this,
we introduce a velocity scaling factor $f_v$, by which we multiply the \citeauthor{canto96} flow speeds before calculating the
line profile. This scaling factor is a free parameter in the fitting described in the following section.

\begin{figure}
\centering
\includegraphics[width=0.9\linewidth, bb=80 50 380 302]{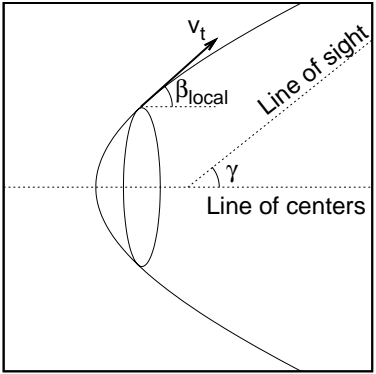}
\caption{Diagram showing the geometry used in the line profile calculations. Each point along the wind-wind collision surface represents a ring of
expanding material, comprising material flowing tangentially along the wind-wind collision surface at speed \vt. Locally, the flow velocity makes an angle
\betalocal\ with the line of centers, while the line of sight makes an angle $\gamma$ with the line of sight. Note that the curvature of the surface
has been exaggerated for clarity.
\label{fig:CantoGeometry}}
\end{figure}

We assume that the wind-wind interaction surface is cylindrically symmetric about the line of centers. Therefore, at each point along
the wind-wind interaction the emission profile is that of an expanding ring. This ring of material flows along the wind-wind collision
surface at speed $\vt = f_v \vC$, where \vC\ is the local flow speed given by the \citet{canto96} equations. Locally, the flow
velocity makes an angle \betalocal\ with the line of centers, as illustrated in Figure~\ref{fig:CantoGeometry}. Note that this is the local shock cone
opening angle, as opposed to asymptotic value which we used in \S\ref{sec:GeometricalModel}. 
We assume that each infinitesimal portion of this ring emits a Dirac $\delta$ function line profile, shifted according to the
line-of-sight velocity $v$. The emission profile $\epsilon(v)$ of the whole ring is then
\begin{equation}
	\epsilon (v) \propto \left[ \vt^2 \sin^2 \betalocal \sin^2 \gamma - (v + \vt \cos \betalocal \cos \gamma)^2 \right]^{-1/2}
\label{eq:epsilon}
\end{equation}
where $\gamma$ is the viewing angle, defined as before as the angle between the line of sight and the line of centers
(see Figs.~\ref{fig:GeometricalModel} and \ref{fig:CantoGeometry}). Note that $\epsilon(v)$ goes to infinity at
$v_\mathrm{min} = \vt (-\sin \betalocal \sin \gamma - cos \betalocal \cos \gamma)$ and 
$v_\mathrm{max} = \vt (\sin \betalocal \sin \gamma - cos \betalocal \cos \gamma)$; $\epsilon(v)$ is undefined outside those velocities.
The function $\epsilon(v)$ goes to infinity because we assume that the intrinsic line profile produced at each point on the ring
is a Dirac $\delta$ function. In reality, the intrinsic line profile produced at each point on the ring will be broadened;
we take this into account in our calculations by convolving the line profile calculated using equation~(\ref{eq:epsilon}) with
a Gaussian (see below). Note also that the integral of $\epsilon(v)$ from $v_\mathrm{min}$ to $v_\mathrm{max}$ is finite,
and is equal to the line luminosity of the expanding ring.

Using this model we cannot calculate the line emissivity at different points along the wind-wind collision surface
self-consistently (unlike, say, the X-ray line model based upon hdyrodynamical
simulations described in \citealp{henley03}). Instead, we adopt a simple formula for calculating the line luminosity
as a function of the distance $x$ measured along the wind-wind collision surface from the stagnation point.
The line luminosity $l(x)$ per unit distance $x$ is given by
\begin{equation}
l(x) = \frac{4}{\pi^{1/2} \xpeak^3} \Lline x^2 \mathrm{e}^{ -( x/\xpeak )^2 },
\label{eq:l(x)}
\end{equation}
where \xpeak\ is the value of $x$ at which $l(x)$ peaks and \Lline\ is the total line luminosity, although in this
model we are only interested in the line \textit{shapes}, so \Lline\ is irrelevant. The form of equation~(\ref{eq:l(x)}) was
chosen after some experimenting with fitting simple functions to the curves in Fig.~2 of \citet{henley03}. The function
$l(x)$ encompasses variations in the temperature, the density, and the
emitting volume per unit $x$. Some examples of $l(x)$ are plotted in Figure~\ref{fig:CantoSurface}. Note that mixing with cooler
material and/or non-equilibrium ionization may affect the form of equation~(\ref{eq:l(x)}), but such effects are beyond the scope
of the present modeling.

Our model line profiles are calculated by summing the individual expanding-ring profiles from each point along the wind-wind collision
surface, weighted by the function $l(x)$. We convolve this summed profile with a Gaussian with $\mathrm{FWHM} = 100~\kmps$ to model thermal
broadening. The resulting profile is then folded through the HETGS response for comparison to the observed profiles, as described below.

\subsection{Comparison to the Observed Profiles}
\label{subsec:ComparisonToObservedProfiles}

The comparison to the observed profiles was carried out using XSPEC\footnote{http://heasarc.gsfc.nasa.gov/docs/xanadu/xspec/} v11.3.2.
We generated a grid of profiles with $\gamma = 5\degr$, 10\degr, 15\degr, ..., 175\degr, $f_v = 1$, 1.25, 1.5, ..., 5, and
$\xpeak = 0.1$, 0.2, 0.4, ..., 6.4. We converted the profiles from velocity space to energy space using the rest energy
of the line we wished to analyze, and used the grid of resulting profiles to generate an XSPEC table model\footnote{http://xspec.gsfc.nasa.gov/docs/xanadu/xspec/xspec11/manual/node61.html}.

In our analysis we concentrated first on the \SiXIV\ \Lyalpha\ line, as the velocity resolution is higher at its wavelength than at
that of the \SXVI\ \Lyalpha\ line, and there are no problems with confusion with nearby lines, unlike the He-like f-i-r triplets.
Our initial approach was to fit the model profiles to the observed lines with $\gamma$, $f_v$ and \xpeak\ all as free parameters.
We also added a power-law component to model the continuum, and for a given observation we fit the model to all four unbinned spectral orders
(HEG~$\pm1$, MEG~$\pm1$) simultaneously, using the $C$~statistic (a modified form of the \citetsq{cash79} statistic, which is implemented in XSPEC).
We applied the model to the first four HETGS observations (CXO$_{001119}$, CXO$_{021016}$, CXO$_{030502}$, and CXO$_{030616}$),
as the last two (CXO$_{030720}$ and CXO$_{030926}$) are contaminated by the CCE component as discussed above.

The best-fitting viewing angles we obtained were similar for all four observations we analyzed: $\approx$34\degr\ for CXO$_{001119}$,
$\approx$22\degr\ for CXO$_{021016}$ and CXO$_{030502}$, and $\approx$14\degr\ for CXO$_{030616}$. This is surprising, given the large
range of phases over which the observations were taken (for example, the phase changed by $\approx$0.1 between CXO$_{021016}$ and CXO$_{030502}$,
yet the best-fitting viewing angles for these two observations differ by $\approx$0.1\degr). We could not find an orbital solution
(specified by $\omega$, $i$, and $e$) which matched the best-fitting viewing angles for all four observations.

\begin{figure*}
\centering
\includegraphics[width=0.96\linewidth]{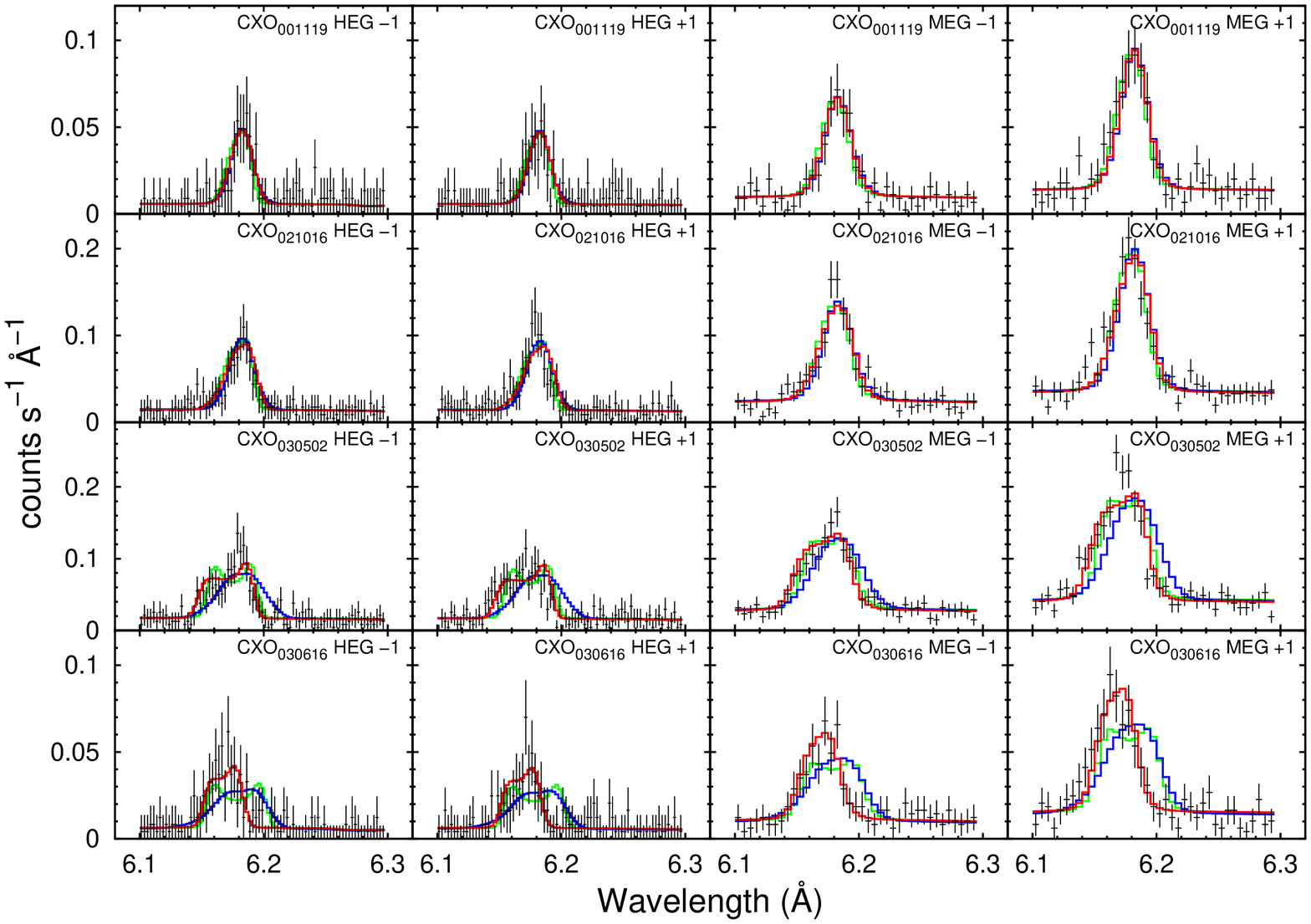}
\caption{Comparison of our model profiles with the observed \SiXIV\ \Lyalpha\ lines, from
CXO$_{001119}$, CXO$_{021016}$, CXO$_{030502}$, and CXO$_{030616}$ (plotted from top to bottom). Each spectral order
is plotted separately. The red curves show the profiles calculated with $\omega = 210\degr$, $i = 70\degr$,
the green curves the profiles calculated with $\omega = 270\degr$, $i = 50\degr$, and the blue curves
the profiles calculated with $\omega = 180\degr$, $i = 50\degr$.\label{fig:SyntheticProfiles}}
\end{figure*}

\begin{figure}
\centering
\includegraphics[width=0.9\linewidth]{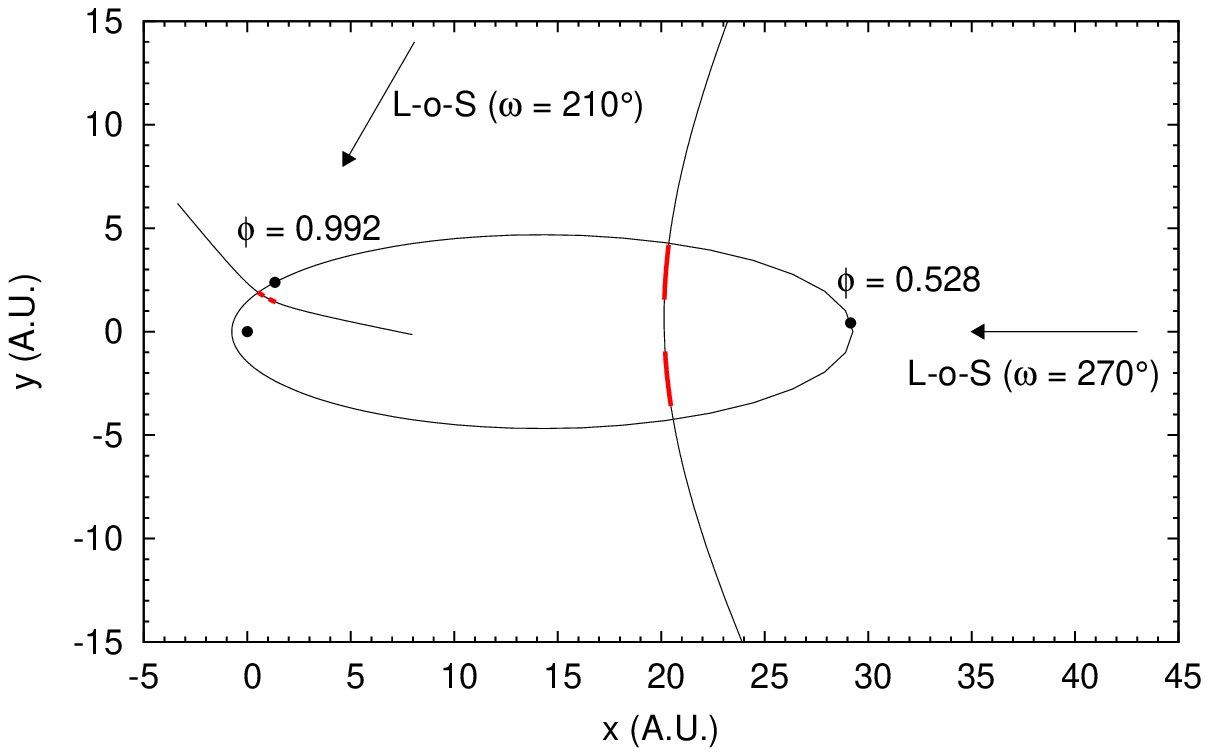}
\caption{The orbit of \ec\ with the eccentricity ($e = 0.95$) from our line profile modeling. All other orbital parameters are the same as those used in Figure~\ref{fig:Orbit}. 
The arrows show the lines of sight (L-of-S) projected into the orbital plane for two different values of the longitude of periastron -- that obtained without
including absorption in the model ($\omega = 210\degr$; see \S\ref{subsec:ComparisonToObservedProfiles}), and that obtained when the effects of absorption are included in
the model ($\omega = 270\degr$; see \S\ref{subsec:Absorption}). Also shown is a comparison of the stellar orientations and geometry of the contact discontinuity for two phases
corresponding to observations CXO$_{001119}$ ($\phi = 0.528$) and CXO$_{030616}$ ($\phi = 0.992$).  In each case the red interval shows the approximate $x$ range where the line
luminosity $l(x)$ (see eq.~[\ref{eq:l(x)}]) is greater than half its maximum value for the \SiXIV\ line. We plot results for the version of the model which includes the effects of absorption
(\S\ref{subsec:Absorption}; lower part of Table~\ref{tab:CantoResults}). The extent of the \SXVI\ emission region is similar to that of the \SiXIV\ emission region.
\label{fig:NewOrbit}}
\end{figure}

We therefore tried a slightly different approach, by trying to find an orbital solution which would give model line profiles consistent with
the observed profiles for all observations. We fixed $e = 0.95$, and for a few sample values of $\omega$ and $i$ we generated theoretical line
profiles and compared them to the observations, allowing \xpeak\ and $f_v$ to vary until the $C$ statistic was minimized. We constrained $f_v$
to be the same for all four observations we investigated. Figure~\ref{fig:SyntheticProfiles} shows these best-fit line profiles for
$\omega = 270\degr$ (green) and $\omega = 180\degr$ (blue), with $i = 50 \degr$ in each case. These values of $\omega$ and $i$ are similar
to the values published by \citet{corcoran01a} and \citet{smith04a}, respectively. They are also similar to the values discussed
in \S\ref{sec:GeometricalModel}. As shown in Figure~\ref{fig:SyntheticProfiles}, these values of $\omega$ and $i$ result
in profiles which have too much emission redward of the \SiXIV\ line center. This is especially true for models in which $\omega=180\degr$ and
$i=50\degr$.  The $\omega=270\degr$, $i=50\degr$ models do a reasonable job in matching the \SiXIV, except for the last observation just before
the start of the X-ray minimum (CXO$_{030616}$). We then attempted to see if we could generate a reasonable fit to all the observed profiles
for some value of $\omega$ and $i$.  After some experimentation, we found that a model with $\omega=210\degr$ and $i=70\degr$ yielded profiles
that provided reasonable descriptions of the shapes of the \SiXIV\ lines in all the observations. These profiles are shown in red in Figure~\ref{fig:SyntheticProfiles}.
The orbit of \ec\ with $e = 0.95$ and $\omega = 210\degr$ is shown in Figure~\ref{fig:NewOrbit} (cf.\ Fig.~\ref{fig:Orbit}).
The best-fitting values of $f_v$ and \xpeak\ are shown in the upper part of Table~\ref{tab:CantoResults}. The \xpeak\ values imply
that the \SiXIV\ emission originates further from the stagnation point (relative to the stellar separation) in the later two
observations, as \xpeak\ is $\sim$8 times larger for these observations. The fact that \xpeak\ is larger
just before periastron than at apastron means that at periastron the \SiXIV\ emission originates from a region with much
higher flow speeds than at apastron (compare the values of \vC\ in Table~\ref{tab:CantoResults}).  This explains
why the model gives relatively narrow lines for CXO$_{001119}$, even though there is material flowing almost directly toward and away
from the observer, and why the model gives lines blueshifted by a few hundred \kmps\ for CXO$_{030616}$, even though the angle between
the flow velocity and the line of sight is large (see Fig.~\ref{fig:NewOrbit}).

We repeated the above fitting with the \SXVI\ \Lyalpha\ line. In general, it could not discriminate between different sets of
orbital parameters as strongly as the \SiXIV\ line, but of those that we investigated, $i = 70\degr$, $\omega = 210\degr$ matched
the observed \SXVI\ profiles the best. Table~\ref{tab:CantoResults} also shows the best-fitting values of $f_v$ and \xpeak\ for \SXVI.
The values of $f_v$ for \SiXIV\ and \SXVI\ are in good agreement. This is as expected, as $f_v$ is a
parameter describing the global flow properties of the wind-wind collision, and so it should not be line dependent.

\begin{deluxetable*}{lccccccccc}
\tabletypesize{\scriptsize}
	\tablewidth{0pt}
\tablecaption{Best-fitting Model Parameters from Synthetic Line Profile Modeling\label{tab:CantoResults}}
\tablehead{
					&
\colhead{\Dsep\tablenotemark{a}}		&
\colhead{\vC\tablenotemark{b}}		&
\multicolumn{3}{c}{\SiXIV}		&&
\multicolumn{3}{c}{\SXVI} \\
\cline{4-6} \cline{8-10}
\colhead{Obs.}				& 
\colhead{(AU)}				&
\colhead{(\kmps)}			& 
\colhead{\xpeak\tablenotemark{c}}	& 
\colhead{$\tau_0$\tablenotemark{d}}	& 
\colhead{$f_v$\tablenotemark{e}}	&&	
\colhead{\xpeak\tablenotemark{c}}	&
\colhead{$\tau_0$\tablenotemark{d}}	&
\colhead{$f_v$\tablenotemark{e}}
}
\startdata
\multicolumn{10}{c}{No absorption -- results for $\omega = 210\degr$, $i = 70\degr$}	\\
\hline
CXO$_{001119}$	& 29.2 & 200	& $0.079^{+0.008}_{-0.007}$ (2.3~AU)	& \nodata			& $2.29^{+0.04}_{-0.02}$	&& $0.119^{+0.026}_{-0.055}$ (3.5~AU)	& \nodata			& $2.25^{+0.05}_{-0.14}$ \\
CXO$_{021016}$	& 17.4 & 340	& $0.140^{+0.008}_{-0.007}$ (2.4~AU)	& \nodata			& 				&& $0.126^{+0.016}_{-0.013}$ (2.2~AU)	& \nodata			& \\
CXO$_{030502}$	& 7.00 & 820	& $0.574^{+0.045}_{-0.041}$ (4.0~AU)	& \nodata			& 				&& $0.302^{+0.043}_{-0.029}$ (2.1~AU)	& \nodata			& \\
CXO$_{030616}$	& 2.73 & 850	& $0.683^{+0.057}_{-0.053}$ (1.9~AU)	& \nodata			& 				&& $0.361^{+0.068}_{-0.041}$ (1.0~AU)	& \nodata			& \\
\hline \\
\multicolumn{10}{c}{Absorption included in model -- results for $\omega = 270\degr$, $i = 50\degr$} \\
\hline
CXO$_{001119}$	& 29.2 & 200	& $0.082^{+0.009}_{-0.008}$ (2.4~AU)	& $0.011^{+0.016}_{-0.011}$	& $3.08^{+0.08}_{-0.17}$	&& $0.127^{+0.023}_{-0.020}$ (3.7~AU)	& $<0.021$			& $2.83^{+0.17}_{-0.12}$ \\
CXO$_{021016}$	& 17.4 & 340	& $0.124^{+0.007}_{-0.006}$ (2.2~AU)	& $0.043^{+0.011}_{-0.012}$	&				&& $0.115^{+0.013}_{-0.011}$ (2.0~AU)	& $0.007^{+0.016}_{-0.007}$	& \\
CXO$_{030502}$	& 7.00 & 820	& $0.143^{+0.007}_{-0.006}$ (1.0~AU)	& $0.192^{+0.014}_{-0.011}$	&				&& $0.127^{+0.012}_{-0.010}$ (0.9~AU)	& $0.165^{+0.024}_{-0.040}$	& \\
CXO$_{030616}$	& 2.73 & 850	& $0.107^{+0.006}_{-0.003}$ (0.3~AU)	& $0.966^{+0.017}_{-0.038}$	&				&& $0.113^{+0.013}_{-0.011}$ (0.3~AU)	& $0.30^{+0.16}_{-0.07}$	& \\
\enddata
\tablenotetext{a}{Stellar separation in AU, using stellar masses from \citet{corcoran01a}, orbital period from \citet{corcoran05b}, and eccentricity $e = 0.95$ from \S\ref{subsec:ComparisonToObservedProfiles}.\\ }
\tablenotetext{b}{Velocity along the contact discontinuity at $x = \xpeak$ for \SiXIV, according to \citet{canto96}.\\ }
\tablenotetext{c}{Distance from the stagnation point at which the line luminosity peaks, in units of the stellar separation and (in parentheses) in AU.\\ }
\tablenotetext{d}{Absorption parameter; see equation~(\ref{eq:tau0}).\\ }
\tablenotetext{e}{Scaling factor by which the velocities calculated from the \citet{canto96} equations are multiplied before calculating model profiles.
                  For each line, the same scaling factor is used for all four observations.\\ }
\end{deluxetable*}

The best-fit values of $f_v$ imply that material is flowing along the wind-wind collision surface at higher speed than is given by the
\citet{canto96} equations. Far from the stagnation point, \vC\ approaches $\sim$900~\kmps. However, the best-fit values of $f_v$
imply that the speed along the collision surface approaches $\sim$2000~\kmps, which is a significant fraction of the
terminal velocity of the companion ($\sim$3000~\kmps; \citealp{pittard02a}). It is expected from hydrodynamical simulations of
the wind-wind collision that the flow speed of the shocked gas approaches the terminal velocity of the companion far from
the stagnation point.

It should be noted that in the above we do not take into account any line-of-sight velocity due to orbital motion.
The observed velocity profile is actually a combination of the projected flow velocity and the orbital velocity
of the line-emitting region. Orbital motion would make the profiles more redshifted if the companion is moving
away from the observer before periastron, and vice versa.	 However, the flow velocity dominates, and we find that we cannot get a reliable estimate of the orbital motion
of the line-emitting region from the data. It should also be noted that even if we could measure the orbital motion
of the line-emitting region, it would not be a direct measure of the orbital motion of the companion. If we could localize the
line-emitting region in space, then we could in principle relate its motion to that of the companion, but in practice this would be difficult to do.

\subsection{The Effect of Bound-Free Absorption on the Observed Line Shapes}
\label{subsec:Absorption}

\citet{henley03} showed that bound-free absorption could have a profound effect on the shapes of the X-ray emission lines observed from colliding wind binaries,
due to differing absorbing columns through the stars' unshocked winds to different parts of the line-emitting region.
When a colliding wind binary is viewed at quadrature, in the absence of any absorption the profiles would be broad, double peaked, and
symmetric about the rest wavelength. However, the absorption of the redshifted emission from the far side of the system can
result in a profile which is strongly positively skewed, with a blueshifted peak and a tail extending to the red.

We have investigated the effect of absorption on our profiles by calculating optical depths through the wind of the companion
star to different points on the wind-wind collision. If we assume the wind is spherically symmetric and non-accelerating, these
optical depths can be calculated analytically \citep{ignace01a}. For portions of the emitting regions which are viewed through the
companion's wind, we can parameterize the strength of the absorption due to the companion's wind with the quantity
\begin{equation}
	\tau_0 = \frac{\kappa \Mdotc}{4 \pi \vc \Dsep},
\label{eq:tau0}
\end{equation}
where \Dsep\ is the stellar separation and $\kappa$ is the opacity at the wavelength of the line of interest.

\begin{figure}
\centering
\includegraphics[width=0.9\linewidth]{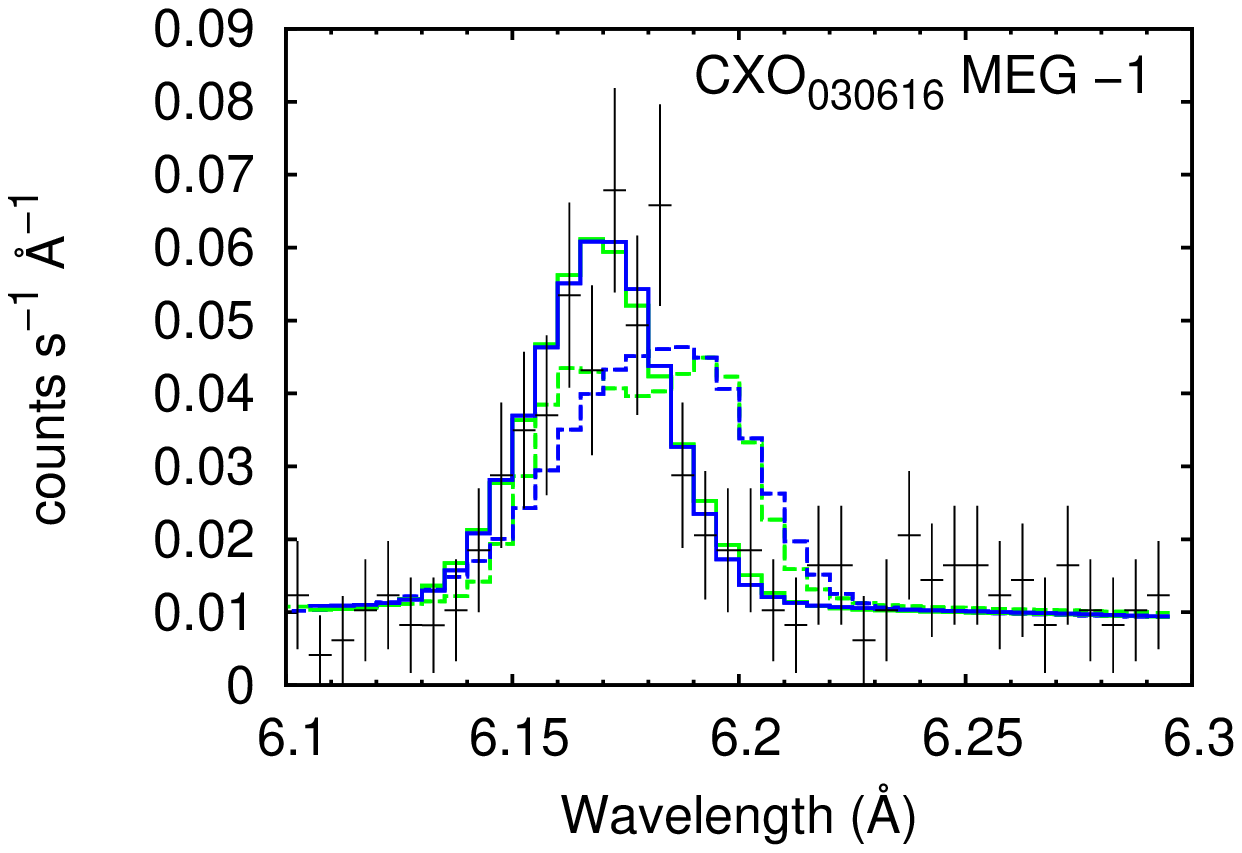}
\caption{Model \SiXIV\ profiles with (\textit{solid}) and without (\textit{dashed}) absorption, compared with the MEG~$-1$ \SiXIV\ line from CXO$_{030616}$.
As in Figure~\ref{fig:SyntheticProfiles}, the green profiles were calculated with $\omega = 270\degr$, $i = 50\degr$, and the blue profiles
with $\omega = 180\degr$, $i = 50\degr$.\label{fig:SyntheticWithAbsorption}}
\end{figure}

We have repeated the fitting described in the previous section, with the addition of $\tau_0$ as a free parameter. With this modification,
we find that we are able to get a good fit (judged by eye) to the \SiXIV\ and \SXVI\ lines for $\omega = 270\degr$, $i = 50\degr$ and 
$\omega = 180\degr$, $i = 50\degr$ (i.e., we do not have to assume a new orientation, as we did in the previous section).
This is shown in Figure~\ref{fig:SyntheticWithAbsorption}, which compares the models with and without absorption for $\omega = 270\degr$ and 180\degr\
with the observed MEG~$-1$ \SiXIV\ line from CXO$_{030616}$ (we choose this observation to illustrate our point because the original
model gave a particularly poor fit to this observaton for $\omega = 270\degr$ and 180\degr).

The lower part of Table~\ref{tab:CantoResults} shows the best-fitting model parameters for the \SiXIV\ and \SXVI\ lines
with absorption included in the model. This set of results is for the orbital orientation $\omega = 270\degr$, $i = 50\degr$.
Note again that the values of $f_v$ for the two lines are in good agreement with each other, as expected. Note also that the
values of \xpeak\ for the two lines are less variable than in the model without absorption, and also that they are in good
agreement with each other. The orbit of \ec\ with $e = 0.95$ and $\omega = 270\degr$ is shown in Figure~\ref{fig:NewOrbit} (cf.\ Fig.~\ref{fig:Orbit}).
Figure~\ref{fig:NewOrbit} also shows the approximate location of the \SiXIV\ emitting region for CXO$_{001119}$ and CXO$_{030616}$.

The results of this modeling are discussed in \S\ref{subsec:ConstrainingInteractionRegion}.

\section{DISCUSSION}
\label{sec:Discussion}

The X-ray emission lines of \ec\ show clear variability, becoming blueshifted and broader just before the X-ray minimum
in mid-2003. This variability cannot be described by a simple geometrical model of the wind-wind collision in which the emission originates from a conical surface with constant opening angle with a longitude of periastron of $\omega\approx270\degr$, which is the value consistent with modeling of the X-ray 2--10~\kev\ flux variations \citep{corcoran01a}  and from analysis of the absorption components in He I P-Cygni features \citep{nielsen07}.  However, we found that a more physical model which describes the shape of the contact surface and the spatial distribution of the X-ray line emissivity along the contact surface, and which takes into account absorption in the unshocked wind of the companion, was able to match the observed HETGS line profiles at all phases with $\omega = 270\degr$ and $i = 50\degr$.

Because of the simplifying assumptions inherent in the line profile model, and the uncertainty of the input parameters, it is possible that other values of $\omega$ and $i$ could be made to fit the observed X-ray line profiles. It should also be emphasized that we did not attempt to find a global best-fitting solution, and so $\omega=270\degr$, $i=50\degr$ is not necessarily the best-fitting set of orbital parameters. Indeed, we found we could also get a good fit to the X-ray line profiles with $\omega = 180\degr$. However, the important result is that we have shown that a colliding wind model can explain the observed line profile variations, without having to invoke any additional flow component \citep{behar07}.

\subsection{Comparison to Line Profiles of Other Massive Stars}

Massive stars are believed to produce X-rays via any or all of the following processes: by wind-wind collisions in binary stars or multiple systems (``colliding wind'' emission); by intrinsic shocks embedded in the unstable, radiatively driven winds (``intrinsic wind'' emission); and via the magnetic field confinement of the radiatively driven wind (``magnetically confined'' emission).  

Stars in which intrinsic wind X-ray emission dominates the observed emission generally show strong line emission but little emission at wavelengths shortward of 3~\AA. 
X-ray emission from a few of these systems have been observed at high spectral resolution. The emission lines are generally broad, with $\mathrm{HWHM} \approx 1000~\kmps$, which is typically half the terminal wind velocity.   \citet{2006ApJ...650.1096L} presented a uniform analysis of the helium-like lines in four O stars ($\zeta$~Ori, $\zeta$~Pup, $\iota$~Ori and $\delta$~Ori) and showed that these stars had rather stronger intercombination lines and lower $\R$ values than we find in the \ec\ spectra; typically the $\R$ ratio was near 2--3 for the \SiXIII\ triplet, while the $\R$ ratio was $1.0\pm0.4$ for \SXV\ in $\zeta$~Pup (the only star for which \SXV\ could be measured).  While $\iota$~Ori and $\zeta$~Ori are binaries, and $\delta$~Ori is a multiple system, none of these four stars show any strong evidence of X-ray emission from wind-wind collisions, and X-rays from all these stars are believed to be dominated by the emission from intrinsic shocks embedded within the winds of the stars themselves.  The $\R$ ratios from these O stars are about a factor of 2 lower than the $\R$ ratios we measured for \ec\ (see Table~\ref{tab:Rratios}), which implies that the minimum radius $r_0$ of the line-forming region for these stars is fairly near the stars, $1.25 < r_0/\rstar < 1.67$ \citep{2006ApJ...650.1096L}, where \rstar\ is the stellar radius. The exception to this is the \SXV\ \R\ ratio from CXO$_{021016}$, which is similar to the value measured from $\zeta$~Pup. An intriguing
possibility is that we are seeing intrinsic emission from the companion's wind. If this is the case, it raises the question of why we see this effect in only one observation.

Colliding wind systems generally show thermal X-ray emission, sometimes at wavelengths shortward of 3~\AA, and may show strong iron K-shell emission.   \citet{pollock05} found that in the long-period, eccentric colliding wind binary WR~140 that the intercombination lines in all the measured He-like lines in that star were very weak, and noted differences between the weak intercombination lines in WR~140 and the stronger $i$ components observed  in O star spectra. The intercombination lines are similarly weak in the shorter-period eccentric colliding wind system $\gamma^{2}$ Velorum \citep{skinner01, henley05a}.

Like the colliding wind systems, magnetically confined winds can show thermal X-ray emission shortward of 3~\AA. The best-studied examples of this class are  $\theta^{1}$~Ori~C and $\tau$ Sco.  Both stars show  hard X-ray emission which is modulated by the rotation of the star, giving rise to variations on much shorter timescales than either the ``intrinsic wind'' emitters or the colliding wind systems.  X-ray emission line profiles from shocked gas in ``magnetically confined'' winds are typically fairly narrow and usually show symmetric profiles.  An analysis of $\theta^{1}$~Ori~C by
\cite{gagne05} showed that the observed emission lines were relatively narrow ($\sim$ few hundred km s$^{-1}$) and that the line centroids are close to zero velocity independent of viewing angle.  This contrasts with the observed variable line profiles we see in \ec.

\subsection{Constraining the Interaction Region}
\label{subsec:ConstrainingInteractionRegion}

The X-ray line profiles offer the most sensitive diagnostic of the flow of the shocked gas produced by the wind-wind interaction in \ec, and one of the few spectral diagnostics which can be unambiguously localized.  We have shown that simple geometrical models in which the interaction region is a conical surface do not do a good job in describing the changes in both line widths and line centroids for any assumed orbital orientation (the longitude of periastron $\omega$), orbital inclination or eccentricity.  A more physical model based on the \cite{canto96} analytical ``thin-shell'' geometry of the wind-wind collision interface provides a reasonable description of the line profile shapes at all of the observed phases for values of eccentricity which are consistent with analyses of the broad-band X-ray fluxes and the \ion{He}{1} P Cygni absorptions. If we do not include absorption in the model, the value of $\omega$ we derive from the line profile modeling (210\degr) is significantly less than the value of $\omega=275\degr$ used in the modeling of the \rxte\ X-ray lightcurve by \citet{corcoran01a}, although it is close to the value of $\omega=200\degr$ derived by \citet{bish01} from his analysis of the \rxte\ data. However, when we take into account the effect of bound-free absorption in the unshocked wind of the companion, we find we can get a good fit to the profiles with $\omega = 270\degr$. We use the model parameters obtained for this latter model (from Table~\ref{tab:CantoResults}) in the following discussion. Our profile analysis is consistent with models in which periastron occurs near superior conjunction, i.e., when the companion star is behind \ec\ as viewed by the observer on earth. This is in contrast to the analysis of \ec's millimeter wavelength flux variations \citep{2005MNRAS.364..922A,2005A&A...437..977A} and the \ion{He}{2} 4686~\AA\ line radial velocity curve \citep{abraham07}, which both concluded that periastron occurs near inferior conjunction.

We first discuss the best-fitting values of $\tau_0$ in Table~\ref{tab:CantoResults}. We can use equation~(\ref{eq:tau0}) to derive the mass-loss rate of the companion
from $\tau_0$. For the opacity we use cross-sections from \citet{balucinska92}, with a revised He cross-section from \citet{yan98}, and solar abundances \citep{anders89}.
Using the stellar separations in Table~\ref{tab:CantoResults}, and assuming $\vc = 3000~\kmps$ for the wind speed of the companion \citep{pittard02a},
the best-fitting values of $\tau_0$ in Table~\ref{tab:CantoResults}
imply mass-loss rates for the companion of $0.6 \times 10^{-5}$~\Msolpy\ for CXO$_{001119}$ to $5.2 \times 10^{-5}$~\Msolpy\ for CXO$_{030616}$
(cf.\ $10^{-5}~\Msolpy$ derived by \citealp{pittard02a} from their modeling of the CXO$_{001119}$ spectrum, using hydrodynamical models of the wind-wind
collsion). Rather than indicating a true variation in the mass-loss rate of the companion, the apparent variation may be a result of the emission region
being smaller at periastron than at apastron. As a result, the lines of sight to the emission region are more likely to pass through the wind acceleration
zone close to the companion, where the densities are higher. These higher densities would enhance the absorption, giving a higher $\tau_0$ in the fitting,
and hence a higher mass-loss rate.

As the absorption cross-section at the energy of the \SXVI\ line is approximately half that at the energy of the \SiXIV\ line \citep{balucinska92}, one would
expect the best-fitting values of $\tau_0$ for \SXVI\ to be approximately half the corresponding values for \SiXIV. Given the uncertainties, our best-fitting
values of $\tau_0$ are consistent with this expectation.

The best-fitting values of \xpeak\ for \SiXIV\ and \SXVI\ are similar to each other, implying they originate from similar regions of the wind-wind collision. As \SiXIV\ and \SXVI\ form over a wide range of temperatures with significant overlap, this result is not surprising. Indeed, models of X-ray line emission in colliding wind binaries based on hydrodynamical simulations show that \SXVI\ and \SiXIV\ are expected to form in similar regions of the wind-wind collision (see Fig.~2 in \citealp{henley03}).

\begin{figure}
\centering
\includegraphics[width=0.9\linewidth]{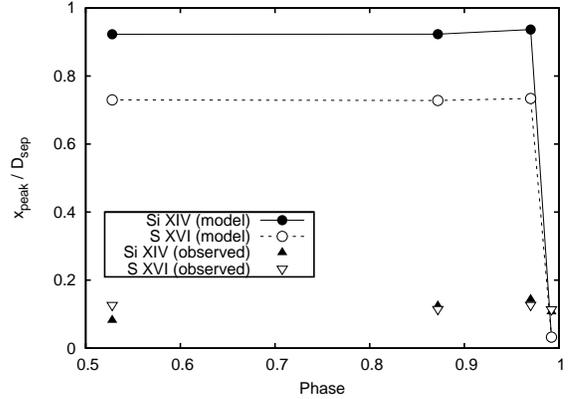}
\caption{The values of \xpeak\ measured from the observed \SiXIV\ and \SXVI\ line profiles (\textit{triangles}), plotted alongside the values of \xpeak\ expected
from hydrodynamical simulations (\textit{circles}). In all cases the values of \xpeak\ have been normalized to the stellar separation.\label{fig:xpeak}}
\end{figure}

We have compared our measured values of \xpeak\ with those expected from hydrodynamical simulations of the wind-wind collision, with stellar wind parameters based on those determined by \citet{pittard02a}. Previous simulations of the colliding winds in $\eta$~Car show that dense clumps from the cold post-shock primary wind can mix into the hot post-shock secondary wind (see Fig.~2 in \citealp{pittard02a}). However, the amount of mixing is dependent on the code and resolution used, and may be reduced by neglected processes, such as magnetic fields. Since the predicted X-ray emission is sensitive to the degree of mixing (J.~M. Pittard, in preparation), we have chosen in this instance to construct a model which minimizes the mixing between the winds. We use the line-profile model of \citet{henley03} to determine the distance from the stagnation point at which the X-ray emission is expected to peak, assuming collisional ionization equilibrium. The values of \xpeak\ expected from these models are plotted in Figure~\ref{fig:xpeak}. Away from periastron ($\phi = 1$), the model values of \xpeak\ are fairly constant. This is as expected if radiative cooling is unimportant, because for adiabatic colliding wind shocks the structure of the wind-wind collision scales self-similarly with binary separation \citep{stevens92}. As a result, we would expect \xpeak\ (in units of the stellar separation, \Dsep) to be constant with orbital phase for a given line. As the system approaches periastron, the increasing post-shock densities make radiative cooling more important, and as a result the size of the emitting region shrinks. Figure~\ref{fig:xpeak} shows that the model values of \xpeak\ rapidly decrease as the system approaches periastron.

Figure~\ref{fig:xpeak} also shows the observed values of \xpeak\ for \SiXIV\ and \SXVI\ from the lower part of Table~\ref{tab:CantoResults}. The observed values of \xpeak\ do not match the values expected from hydrodynamical simulations. Away from periastron the observed values are $\sim$7--10 times smaller than the model values. Also, the observed value of \xpeak\ does not rapidly fall just before periastron. Instead, \xpeak\ is fairly constant at $\sim$0.1. Therefore, away from periastron the observed lines are forming closer to the stagnation point than is expected from the hydrodynamical simulations. This may be because cooling in the shocked gas is more rapid than pure radiative cooling, perhaps due to inverse Compton cooling or mixing with cool material. This increased cooling would result in a smaller emission region than would be expected if only radiative cooling were operating. However, this enhanced cooling is unlikely to be more important at apastron, when the densities are lower, than at periastron -- just before periastron ($\phi = 0.992$; CXO$_{030616}$), the observed and model values of \xpeak\ are in better agreement. Furthermore, significant mixing would lead to velocities along the wind-wind collision surface approaching the analytical model of \citet{canto96}, in which complete mixing is assumed, whereas we observe velocities $\sim$2--3 times larger than the \citeauthor{canto96} values. An alternative explanation is that the shocked gas is out of equilibrium. The lower post-shock densities away from apastron may mean that the ionization temperature lags behind the kinetic temperature after the gas has been shock heated. As a result, the \SiXIV\ and \SXVI\ emission would tend to originate closer to the stagnation point than would be expected if the gas were in equilibrium -- if the gas were in equilibrium, it would be too hot near the stagnation point for \SiXIV\ and \SXVI. As the system approaches periastron, the post-shock densities increase, and the shocked gas starts to equilibrate. By itself, this equilibration would tend to increase the size of the emission region in units of \Dsep, as the gas near the stagnation point becomes too hot for \SiXIV\ and \SXVI. However, as the system approaches periastron, radiative cooling becomes more important, which tends to reduce the size of the emission region in units of \Dsep. Our observed values of \xpeak\ suggest that, as the system approaches periastron, the increase in the size of the emission region due to the gas equilibrating is canceled by the decrease in the size of the emission region due to radiative cooling, resulting in a roughly constant observed value of \xpeak. More detailed hydrodynamical modeling, including modeling of the post-shock ionization evolution, is required to see if the above-described scenario is likely. Such modeling is beyond the scope of this paper. However, we can in principle estimate whether or not non-equilibrium ionization is likely to be important by comparing the ionization timescale to the flow timescale, using equation~(14) from \citet{henley05a}. We find the ratio of the ionization timescale to the flow timescale is 0.94 for CXO$_{001119}$, 0.56 for CXO$_{021016}$, and 0.23 for CXO$_{030502}$. Unfortunately, these results are inconclusive: as the ratios are neither much greater than nor much less than unity, it is difficult to state with certainty whether or not non-equilibrium effects are expected in these observations. As radiative cooling is more important for CXO$_{030616}$, the appropriate ratio to calculate is that of the ionization timescale to the radiative timescale (eq.~[15] from \citealp{henley05a}), which is 0.01. As this ratio is much less than unity, this implies that the system is in equilibrium just before periastron, as we have suggested above.

The fact that \xpeak\ (in units of \Dsep) is roughly constant implies that strong cooling is not affecting the \SiXIV\ and \SXVI\ emission as the system approaches periastron -- if it were, the emitting region would decrease in size relative to the stellar separation, as regions further from the stagnation point would become too cool to emit. In contrast to this, \citet{hamaguchi07} and Paper II present evidence of cooling in the X-ray--emitting plasma based on excess emission on the low energy side of the \ion{Fe}{25} triplet near periastron, caused by rapid cooling driving the hot gas out of equilibrium as the stars approach one another. However, it is possible that while the high densities near the stagnation point result in strong cooling in that region, where the \FeXXV\ emission originates, cooling is less important further out, where the \SXVI\ and \SiXIV\ emission originates. Even if the shocked material is cooling rapidly near the stagnation point, it is still possible to have material hot enough to emit \SXVI\ and \SiXIV\ further out, because material is being shock-heated all along the shock surface.

If the shocked gas is in equilibrium, then the innermost region, nearest the apex of the shock cone, is too hot to emit much silicon or sulfur line emission when there is not strong cooling, mixing from clumps of post-shock primary wind, or clumps within the secondary wind. We take the inner region of the shock cone where the \SXVI\ emissivity is less than 50\%\ of its maximum emissivity to be the region where the temperature is high enough that sulfur is almost completely ionized, and assume that the \FeXXV\ and \FeXXVI\ emission originates from somewhere in this region.  This assumption constrains the iron K-shell region to be on the surface of the wind-wind interface within about 0.14~AU just before periastron. If, as discussed above, the shocked gas is not in equilibrium in the earlier observations, we cannot use this method to constrain the size of the iron K-shell region near apastron.

\section{CONCLUSIONS}
\label{sec:Conclusions}

We have presented our analysis of resolved silicon and sulfur X-ray emission lines from a series of HETGS observations of \ec\ at key orbital phases.  These lines originate in the wind-wind collision zone where the slow, dense wind of \ec\ interacts with the fast, low-density wind of a massive companion star.  We have shown that line profile variations around the orbit are not consistent with simple geometrical models of the line forming region.  A more physically realistic model which takes into account the detailed geometry of the contact discontinuity and allows for variations in the emissivity distribution along the shock boundary can produce both the variations in the line centroids and the observed changes in line width. This analysis allows us to probe directly both the temperature distribution along the shock boundary and also the flow of the shocked wind of the companion away from the stagnation point at the apex of the shock.

The \SiXIV\ and \SXVI\ emission appear to originate from similar regions, which is as expected given the range of temperatures at which they are produced. However, away from periastron they originate closer to the stagnation point than is expected from hydrodynamical simulations. This may be because the wind-wind collision is out of equilibrium, and the line emission is originating from a region which would be too hot if the wind-wind collision were in equilibrium (although it should be noted that we did not find unambiguous evidence of non-equilibrium conditions from the line flux ratios). Just before periastron the size of the \SiXIV-emitting region is closer to that which is expected from a hydrodynamical simulation, suggesting that the shocked gas has equilibrated at the time of CXO$_{030616}$. We find that the flow speed along the wind-wind collision surface is $\sim$3$\times$ the flow speed given by the analytic model of \citet{canto96}. This larger flow speed approaches a large fraction of the terminal velocity of the companion star's wind far from the stagnation point, which is in fair agreement with detailed numerical hydrodynamic models of the flow, and suggests that the mixing of the cold postshock primary wind into the hot postshock primary wind is relatively unimportant, perhaps due to the presence of magnetic fields.

We can obtain a good fit to the profiles with an inclination and longitude of periastron similar to those which have previously been assumed ($i \approx 50\degr$ and $\omega \approx 270\degr$), although this is only true if we include the effects of absorption by the unshocked wind of the companion. Given the simplifying assumptions inherent in the model, the uncertainty of the various input model parameters, and the fact that we did not search for a global best fit, we cannot rule out other possible orbital orientations. Nevertheless, an important result of this analysis is that it shows that colliding wind models can fit the detailed flow dynamics as shown by the variations in X-ray line profiles, without recourse to additional flows in the system.

These results must be considered preliminary since the observed line profile variations need to be confirmed as dependent on orbital phase rather than simply secular changes in the wind. \chandra\ HETGS observations are scheduled for Fall 2008 and late 2008, and ideally more will be carried out around the time of the next X-ray minimum, expected in 2009 January. These observations will complement ongoing X-ray monitoring with \rxte\ and ground-based optical and radio observations. Meanwhile, the line profile model can be improved by the inclusion of more realistic absorption from the wind of \ec\ and from the wind of the companion star, which might give further insight into the mass loss rate of the wind from the companion, and by more detailed numerical models including turbulence to directly determine the dependence of the theoretical line profiles on the orientation and geometry of the colliding wind region.  

\acknowledgements

We gratefully acknowledge the exceptional support of
Dr.\ Fred Seward and Dr.\ Norbert Schulz of the Chandra X-ray
Center for their help in scheduling these observations.
We would also like to thank Dr.\ John Hillier and the referee,
Dr.\ Maurice Leutenegger, whose comments have greatly
improved the manuscript.
This work was supported by SAO grant GO3-4008A.
This research has made use of NASA's Astrophysics Data
System. This research has made use of software obtained from
the High Energy Astrophysics Science Archive Research
Center (HEASARC), provided by NASA's Goddard Space
Flight Center, and software developed and provided by the \chandra\ X-ray Center. 
DBH gratefully acknowledges funding from the School of Physics and Astronomy
at the University of Birmingham, and from NASA grant NAG5-NNGO4GD78G.
JMP gratefully acknowledges funding from the Royal Society.

\bibliography{references}

\end{document}